\documentclass[12pt,preprint]{aastex}

\shorttitle{RAISHIN: New GRMHD Code}
\shortauthors{Mizuno et al.}

\usepackage{graphicx}

\begin{document}

\title{RAISHIN: A High-Resolution Three-Dimensional General \\
Relativistic Magnetohydrodynamics Code}

\author{
Yosuke Mizuno\altaffilmark{1} \altaffilmark{6}, Ken-Ichi
Nishikawa\altaffilmark{1} \altaffilmark{2}, Shinji
Koide\altaffilmark{3}, Philip Hardee\altaffilmark{4} \\ and Gerald 
J. Fishman\altaffilmark{5} }

\altaffiltext{1}{National Space Science and Technology Center, 320
Sparkman Drive, VP 62, Huntsville, AL 35805, USA;
Yosuke.Mizuno@msfc.nasa.gov} \altaffiltext{2}{Center for Space
Plasma and Aeronomic Research, University of Alabama in Huntsville}
\altaffiltext{3}{Department of Physics, Kumamoto University,
Kurokami, Kumamoto, 860-8555, Japan} \altaffiltext{4}{Department of
Physics and Astronomy, The University of Alabama, Tuscaloosa, AL
35487, USA} \altaffiltext{5}{NASA-Marshall Space Flight Center,
National Space Science and Technology Center, 320 Sparkman Drive, VP
62,Huntsville, AL 35805, USA} \altaffiltext{6}{NASA Postdoctoral
Program Fellow/ NASA Marshall Space Flight Center}

\shorttitle{RAISHIN: A Code for 3D GRMHD}

\begin{abstract}

We have developed a new three-dimensional general relativistic
magnetohydrodynamic (GRMHD) code, RAISHIN, using a conservative,
high resolution shock-capturing scheme. The numerical fluxes are
calculated using the Harten, Lax, \& van Leer (HLL) approximate
Riemann solver scheme. The flux-interpolated, constrained transport
scheme is used to maintain a divergence-free magnetic field. In
order to examine the numerical accuracy and the numerical
efficiency, the code uses four different reconstruction methods:
piecewise linear methods with Minmod and MC slope-limiter function,
convex essentially non-oscillatory (CENO) method, and piecewise
parabolic method (PPM) using multistep TVD Runge-Kutta time advance
methods with second and third-order time accuracy. We describe code
performance on an extensive set of test problems in both special and
general relativity. Our new GRMHD code has proven to be accurate in
second order and has successfully passed with all tests performed,
including highly relativistic and magnetized cases in both special
and general relativity.

\end{abstract}
\keywords{accretion, accretion disks - black hole physics -
magnetohydrodynamics: (MHD) - method: numerical -relativity}

\section{Introduction}

Both magnetic and gravitational fields play an important role in
determining the evolution of the matter in many astrophysical
objects. In highly conducting plasma, the magnetic field can be
amplified by gas contraction or shear motion. Even when the magnetic
field is weak initially, the magnetic field grows short time scales
and influences the gas dynamics of the system. This is particulary
important for a compact object such as a black hole or a neutron
star. Relativistic jets have been observed or postulated in various
astrophysical objects, including active galactic nuclei (AGNs)
(e.g., Urry \& Pavovani 1995; Ferrari 1998; Blandford 2002),
microquasars in our galaxy (e.g., Mirabel \& Rodr\'{i}guez 1999), and
gamma-ray bursts (GRBs) (e.g., Zhang \& M\'{e}sz\'{a}ros 2004; Piran
2005; M\'{e}sz\'{a}ros 2006). The most promising mechanisms for
producing the relativistic jets involve the magnetohydrodynamic
centrifugal acceleration  and/or magnetic pressure driven
acceleration from the accretion disk with the compact objects (e.g.,
Blandford \& Payne 1982; Fukue 1990), or the extraction of rotating
energy from the rotating black hole (Penrose 1969; Blandford \&
Znajek 1977). In addition, the differential rotation of the plasma
in the disk raises the magnetorotational instability (MRI), which
plays an important role for the transportation of angular momentum
from the disk due to the associated turbulence in accretion disks
(Balbus \& Hawley 1991, 1998). It is much more efficient than the
internal disk transport. Magnetars are neutron stars with extremely 
large magnetic fields ($\sim 10^{14}-10^{15} G$), as inferred from 
studies of anomalous X-ray pulsars and soft gamma-ray repeaters 
(e.g., Woods \& Thompson 2006)). These intense magnetic fields are 
expected to affect the internal structure of the neutron star 
(e.g., Bocquet et al. 1995). The less intense fields of other neutron stars, 
 $\sim 10^{12} - 10^{13} G$, may also affect the internal structure.

There has been much progresses in the analytical studies of
relativistic astrophysical phenomena (e.g., Camenzind 1990; Fendt
1997; Fendt \& Ouyed 2004). However, the problem is complex,
involving time-dependent, three-dimensional dynamics of magnetized
plasmas in the relativistic potential. Therefore analytical
solutions are rather limited to simplified cases that are time
stationary and spatially symmetric. Because of these, numerical
experiments play an important role for complimenting theoretical
works.

A complete review of numerical approaches to relativistic
hydrodynamics is given by Mart\'{i} \& M\"{u}ller (2003) and Font
(2003). Numerical codes in special relativistic magnetohydrodynamics
(SRMHD) have been developed by a growing number of authors (van
Putten 1993; Koide et al. 1996; Komissarov 1999a;
Balsara 2001; Koldoba et al. 2002; Del Zanna,
Buccianti, \& Londrillo 2003; Leismann et al. 2005; Mignone \& Bodo
2006) and have been applied to the study of the relativistic jets
(Koide et al. 1996, 1997; Koide 1997;
Nishikawa et al. 1997, 1998; Kommisarov 1999b; Leismann et al. 2005;
Mignone \& Bodo 2006) and pulsar wind nebulae (Kommisarov \&
Lyubarsky 2003, 2004; Bucciantini et al. 2003, 2004, 2005, 2006; 
Del Zanna et al. 2004).
In addition, Komissarov (1999a) and Balsara (2001) proposed a
comprehensive set of tests to validate numerical codes in special
relativity. The exact solutions of the Riemann problem in the SRMHD
have been obtained by Giacomazzo \& Rezzolla (2006).

In order to investigate relativistic magnetorotators (RMRs), general
relativistic magnetohydrodynamics (GRMHD) codes with fixed
spacetimes have been developed by some authors (Yokosawa 1993;
Koide et al.1998; De Villiers \& Hawley 2003; Gammie et al. 2003; 
Komissarov 2004; Ant\'{o}n et al. 2005;
Anninos et al. 2005).  These codes have been used to
study the structure of accretion flows onto Kerr black holes and/or
the formation of jets (Yokosawa 1995; Koide et al. 1998,
1999, 2000; Nishikawa et al. 2005; De Villiers et al. 2003, 2005a; 
Hirose et al. 2004; Krolik et al. 2005; Hawley \& Krolik 2006; 
McKinney \& Gammie 2004; McKinney 2005, 2006), the Blandford-Znajek effect near the
rotating black hole (Koide et al. 2002; Koide 2003; Komissarov
2005), and the formation of GRB jets in collapsars (Mizuno et al.
2004a, 2004b; De Villiers et al. 2005b).

Few attempts have been made to simulate relativistic MHD flows in
dynamical spacetimes (e.g., Wilson 1975, Baumgarte \& Shapiro 2003).
Recently new codes capable of evolving the Einstein-Maxwell-MHD
equations have been developed by Duez et al. (2005), Shibata \&
Sekiguchi (2005) and Anderson et al. (2006).
These codes have been applied to investigate
rotating neutron stars (Duez et al. 2006a,b), and collapsing neutron
stars for a central engine for short gamma-ray bursts (Shibata et
al. 2006).  These new  GRMHD simulations with dynamical spacetimes
will be applied to investigate various astrophysical systems such as
merging neutron stars and black holes including gravitational waves
and  relativistic jets (e.g., Duez et al. 2005).

Our previous GRMHD code developed by Koide has been applied to many
high-energy astrophysical phenomena and showed pioneering results
(Koide et al. 1998, 1999, 2000, 2002; Koide 2003, 2004, 2006; 
Mizuno et al. 2004a,b; Nishikawa et al. 2005). 
However, the code cannot perform calculations in highly
relativistic or highly magnetized regimes. The critical problem of
the previous GRMHD code is that it cannot guarantee a
divergence-free magnetic field. Even though we introduce a
divergence-cleaning step in the code, it cannot perform long-term
evolution, except in special cases. In order to overcome these
numerical difficulties, we have developed a new, three-dimensional,
GRMHD code called RAISHIN, for RelAtIviStic magnetoHydrodynamic
sImulatioN. (RAISHIN is the ancient Japanese god of lightning.) It
uses a conservative, high resolution shock-capturing scheme.

The structure of the paper is as follows. In section 2 we describe the
basic equations of GRMHD including the essentials of the 3$+$1
formalism, the description of the magnetic field, the induction
equation and the conservation equations of particle number, and the
stress-energy tensor in conservative form. In section 3 we describe
the basic algorithm of our new GRMHD code based on the recent modern
techniques such as HLL approximate Riemann solver, various
reconstruction methods and flux-interpolated constrained transport.
We show the performance of the code on a series of test problems in
section 4. The summary and conclusions are given in section 5.

The new simulation results of jet formation using this code will be
reported separately (Mizuno et al. 2006).

\section{Basic Equations}

\subsection{Spacetimes and observers}

We investigate the evolution of a magnetized fluid in the
background spacetime metric of a black hole written as
\begin{equation}
ds^{2}= - \alpha^{2} dt^{2} + \gamma_{ij} (dx^{i}+ \beta^{i}dt)
(dx^{j}+\beta^{j}dt),
\label{eq2-1}
\end{equation}
where $\alpha$, $\beta^{i}$, and $\gamma_{ij}$ are the lapse
function, shift vector, and spatial metric respectively. A natural
observer associated in the metric given by Eq. (\ref{eq2-1}) is the
so-called Eulerian observer with four-velocity $u^{\mu}$
perpendicular to the hypersurfaces of constant $t$ at each event in
the spacetime. The covariant and contravariant components of
$n^{\mu}$ are given by
\begin{equation}
n^{\mu} = {1 \over \alpha}(1, - \beta^{i}), \label{eq2-2}
\end{equation}
and
\begin{equation}
n_{\mu} = (-\alpha, 0, 0, 0 ), \label{eq2-3}
\end{equation}
respectively.

The comoving observer follows the fluid motion with four-velocity $u^{\mu}$.
The three-velocity of the fluid as measured by the Eulerian
observer can be given as
\begin{equation}
v^{i} \equiv -{h^{i}_{j} u^{\mu} \over n_{\mu} u^{\mu}}, \label{eq2-4}
\end{equation}
where $- n_{\mu} u^{\mu} \equiv W $ is the relative Lorentz factor
between $u^{\mu}$ and $n_{\mu}$ and $h_{\mu \nu} = g_{\mu \nu} +
n_{\mu} n_{\nu}$ is the projector tensor onto the hypersurface
orthogonal to $n^{\mu}$. The spatial component of the projector
tensor is $h_{ij}=\gamma_{ij}$. Eq (\ref{eq2-4}) is written as
\begin{equation}
v^{i} = {u^{i} \over \alpha u^{t}} + {\beta^{i} \over \alpha},
\label{eq2-5}
\end{equation}
and
\begin{equation}
v_{i} = u_{i} /\gamma. \label{eq2-6}
\end{equation}
The Lorentz factor satisfies the following relation
\begin{equation}
\gamma = {1 \over \sqrt{(1- v^{2})}} = \alpha u^{t}, \label{eq2-7}
\end{equation}
where $v^{2}=\gamma_{ij}v^{i}v^{j}$.

\subsection{Evolution of the electromagnetic fields}

The electromagnetic field in general relativity is described by the
Faraday electromagnetic tensor $F^{\mu \nu}$. This tensor is related
to the electric field $E^{\mu}$ and magnetic field $B^{\mu}$
measured by the Eulerian observer,
\begin{equation}
F^{\mu \nu} = n^{\mu} E^{\nu} - n^{\nu} E^{\mu} + n_{\gamma}
\epsilon^{\gamma \mu \nu \sigma} B_{\sigma}, \label{eq2-8}
\end{equation}
where $\epsilon^{\gamma \mu \nu \sigma} = (-g)^{1/2} [\gamma \mu \nu
\sigma]$, $g$ is the determinant of the four-metric ($g = det g_{\mu
\nu}$) and $\epsilon^{\gamma \mu \nu \sigma}$ is the antisymmetric
Levi-Civita symbol. Both electric and magnetic fields are orthogonal
to $n^{\mu}$ ($E^{\mu} n_{\mu}=B^{\mu}n_{\mu}=0$), and can be
expressed as
\begin{equation}
E^{\mu} = F^{\mu \nu}n_{\nu},  \label{eq2-9}
\end{equation}
and
\begin{equation}
B^{\mu} = {1 \over 2} \epsilon^{\mu \nu \kappa \lambda} n_{\nu} F_{\kappa \lambda} =
n_{\nu} F^{* \nu \mu}, \label{eq2-10}
\end{equation}
where
\begin{equation}
F^{* \mu \nu} = {1 \over 2} \epsilon^{\mu \nu \kappa \lambda} F_{\kappa \lambda}
\label{eq2-11}
\end{equation}
 is dual of the electromagnetic field tensor.

 We adopt the ideal MHD condition and assume that the fluid is
 a perfect conductor. In this case the fluid has infinite
 conductivity and in order to keep the current finite, the
 conduction current must vanish
\begin{equation}
 F^{\mu \nu} u_{\nu} = 0 \label{eq2-12}
\end{equation}
which means that the electric field in the rest frame of the
fluid is zero. The electric and magnetic fields measured by
a comoving observer are
\begin{equation}
\hat{E}^{\mu} = F^{\mu \nu} u_{\nu}, \label{eq2-13}
\end{equation}
and
\begin{equation}
\hat{B}^{\mu} = u_{\nu} F^{* \nu \mu}. \label{eq2-14}
\end{equation}
The ideal MHD condition (\ref{eq2-12}) satisfies the condition
that the electric field observed by a comoving observer becomes zero
($\hat{E}^{\mu} = 0$). $\hat{B}^{\mu}$ is orthogonal to $u_{\mu}$,
i.e. $u_{\mu}\hat{B}^{\mu}=0$. The electromagnetic tensor can be
expressed by the terms of $\hat{B}^{\mu}$ as
\begin{equation}
F^{\mu \nu} = u_{\gamma} \epsilon^{\gamma \mu \nu \sigma} \hat{B}_{\sigma}.
\label{eq2-15}
\end{equation}
The dual of the electromagnetic field tensor is also obtained by
\begin{equation}
F^{* \mu \nu} = \hat{B}^{\mu} u^{\nu} - \hat{B}^{\nu} u^{\mu}. \label{eq2-16}
\end{equation}
Since $\hat{B}^{\mu}$ is orthogonal to $u_{\mu}$, we have
$h^{\mu}_{\nu} \hat{B}^{\nu} =
\hat{B}^{\mu}$.
From Eqs. (\ref{eq2-10}) and (\ref{eq2-16})
\begin{equation}
h^{\mu}_{\nu} B^{\nu} = - n_{\lambda}u^{\lambda} \hat{B}^{\mu}. \label{eq2-17}
\end{equation}
Therefore
\begin{equation}
\hat{B}^{\mu} = -{h^{\mu}_{\nu} B^{\nu} \over n_{\nu}u^{\nu}}. \label{eq2-18}
\end{equation}
The time and space components of Eq. (\ref{eq2-18}) are given by
\begin{equation}
\hat{B}^{t} = { \gamma v_{i}B^{i} \over \alpha }, \label{eq2-19}
\end{equation}
and
\begin{equation}
\hat{B}^{i} = {B^{i} + \alpha \hat{B}^{t}u^{i} \over \alpha u^{t}}. \label{eq2-20}
\end{equation}

The evolution equation for the magnetic field can be obtained in
conservation form from the dual of Maxwell's equation
\begin{equation}
F_{\mu \nu, \lambda} + F_{\lambda \mu, \nu} + F_{\nu \lambda, \mu} =0
\end{equation}
in a coordinate basis,
\begin{equation}
F^{* \mu \nu}_{; \nu} = {1 \over \sqrt{-g}} {\partial \over \partial x^{\nu}}
(\sqrt{-g} F^{* \mu \nu}) =0. \label{eq2-21}
\end{equation}
Since $\sqrt{-g}= \alpha \sqrt{\gamma}$ and $F^{* it} = B^{i}/ \alpha$,
the time component of Eq. (\ref{eq2-21}) gives the divergence-free
magnetic field constraint
\begin{equation}
{1 \over \sqrt{-g}}{\partial \over \partial x^{i}} (\sqrt{\gamma} B^{i} ) = 0,
\label{eq2-22}
\end{equation}
 and the spatial components of Eq. (\ref{eq2-21}) gives the induction equation
\begin{equation}
{1 \over \sqrt{-g}}{\partial \over \partial t} (\sqrt{\gamma} B^{i}) + {1 \over
\sqrt{-g}}{\partial \over \partial x^{i}} [ \sqrt{-g}(u^{j} \hat{B}^{i} - u^{i}
\hat{B}^{j})] = 0. \label{eq2-23}
\end{equation}
It follows from Eq. (\ref{eq2-20}) that
\begin{equation}
u^{j}\hat{B}^{i} - u^{i} \hat{B}^{j} = (\tilde{v}^{j} B^{i} - \tilde{v}^{i} B^{j}),
\label{eq2-24}
\end{equation}
where $\tilde{v}^{i} = v^{i} - \beta^{i}/\alpha$. Therefore the
induction equation can be written as
\begin{equation}
{1 \over \sqrt{-g}}{\partial \over \partial t} (\sqrt{\gamma} B^{i}) + {1 \over
\sqrt{-g}}{\partial \over \partial x^{i}} [ \sqrt{-g}(\tilde{v}^{j} B^{i} -
\tilde{v}^{i} B^{j})] = 0. \label{eq2-25}
\end{equation}

\subsection{Conservation Equations}

The evolution equations for matter can be expressed as the local
conservation laws for particle number and energy-momentum. The
particle number conservation equation is written as
\begin{equation}
(\rho u^{\mu})_{; \mu}=0. \label{eq2-26}
\end{equation}
 Here $\rho$ is the rest-mass density. In a coordinate basis we rewrite this as
\begin{equation}
 {1 \over \sqrt{-g}}{\partial \over \partial x^{\mu}} (\sqrt{-g} \rho u^{\mu}) =0,
 \label{eq2-27}
\end{equation}
and in 3+1 formalism
\begin{equation}
{1 \over \sqrt{-g}} {\partial \over \partial t} (\sqrt{\gamma} D) + {1 \over \sqrt{-g}}
{\partial \over \partial x^{i}} (\sqrt{-g} D \tilde{v}^{i}) =0, \label{eq2-28}
\end{equation}
where $D= \gamma \rho$.

The energy-momentum conservation equation is given by
\begin{equation}
T^{\mu \nu}_{; \nu} =0, \label{eq2-29}
\end{equation}
where $T^{\mu \nu}$ is the energy-momentum tensor. In a coordinate
basis we rewrite this as
\begin{equation}
{1 \over \sqrt{-g}} {\partial \over \partial x^{\nu}} (\sqrt{-g} T^{\mu \nu}) =0,
\label{eq2-30}
\end{equation}
and in 3+1 formalism
\begin{equation}
{1 \over \sqrt{-g}} {\partial \over \partial t} (\sqrt{-g} T^{t \nu}) + {1 \over
\sqrt{-g}} {\partial \over \partial x^{i}} (\sqrt{-g} T^{i \nu}) - \Gamma^{\mu}_{\sigma
\nu} T^{\sigma \nu} =0, \label{eq2-31}
\end{equation}
where $\Gamma^{\mu}_{\sigma \nu}$ is the Christoffel symbol. The
energy-momentum tensor for a system containing a perfect fluid and
an electromagnetic field is the sum of a fluid part,
\begin{equation}
T^{\mu \nu}_{\rm{fluid}} = \rho h u^{\mu} u^{\nu} + p g^{\mu \nu}, \label{eq2-32}
\end{equation}
where $p$ is pressure, $h$ is the specific enthalpy, defined by $h =
1 + u +p / \rho$ and $u$ is internal energy and an electromagnetic
part,
\begin{equation}
T^{\mu \nu}_{\rm{EM}} = {1 \over 4 \pi} \left( F^{\mu \lambda} F^{\nu}_{\lambda} -
{1 \over 4} g^{\mu \nu} F_{\alpha \beta} F^{\alpha \beta} \right) . \label{eq2-33}
\end{equation}
In the ideal MHD condition, $T^{\mu \nu}_{\rm{EM}}$ can be expressed
simply in term of a magnetic 4-vector $b^{\mu}= \hat{B}^{\mu}/\sqrt{4 \pi}$
as (e.g., Eq. (\ref{eq2-15}))
\begin{equation}
T^{\mu \nu}_{\rm{EM}} = b^{2} u^{\mu} u^{\nu} +  {b^{2} \over 2}
g^{\mu \nu} - b^{\mu} b^{\nu}, \label{eq2-34}
\end{equation}
where $b^{2}= b^{\nu}b_{\nu}$.
Hence the energy-momentum tensor is given by
\begin{equation}
T^{\mu \nu} = (\rho h + b^{2}) u^{\mu} u^{\nu} + \left( p + {b^{2} \over 2} \right)
g^{\mu \nu} - b^{\mu} b^{\nu}. \label{eq2-35}
\end{equation}
The spatial components of the energy-momentum conservation equation
give the momentum equation
\begin{equation}
{1 \over \sqrt{-g}} {\partial \over \partial t} (\sqrt{\gamma} S_{i}) + {1 \over
\sqrt{-g}} {\partial \over \partial x^{i}} (\sqrt{-g} T^{j}_{i}) = T^{\mu \nu}
\left( {\partial g_{\nu i} \over \partial x^{\mu}} - \Gamma^{\sigma}_{\nu \mu}
g_{\sigma i} \right), \label{eq2-36}
\end{equation}
where $S_{i}$ is the momentum density of the magnetized fluid
\begin{equation}
S_{i} = \alpha T^{t}_{i} = (\rho h + b^{2}) \gamma^{2} v_{j} - \alpha b^{t} b_{j}.
\label{eq2-37}
\end{equation}
The time component of the energy-momentum conservation equation
gives the energy equation
\begin{equation}
{1 \over \sqrt{-g}} {\partial \over \partial t} (\sqrt{\gamma} \tau) + {1 \over
\sqrt{-g}} {\partial \over \partial x^{i}} [\sqrt{-g} (\alpha T^{ti} - D
\tilde{v}^{i})] = \alpha \left( T^{\mu t} {\partial \ln \alpha \over \partial x^{\mu}}
- T^{\mu \nu} \Gamma^{t}_{\nu \mu} \right), \label{eq2-38}
\end{equation}
where $\tau$ is the total energy density
\begin{equation}
\tau = \alpha^{2} T^{tt} - D = (\rho h + b^{2}) \gamma^{2} - (p + b^{2}/2) -
\alpha^{2} (b^{t})^{2} - D . \label{eq2-39}
\end{equation}

To complete the system of equations, it remains to specify the
equation of state (EOS). In this paper we adopt a $\Gamma$-law EOS
\begin{equation}
p = (\Gamma- 1) \rho u, \label{eq2-40}
\end{equation}
where $\Gamma$ is adiabatic index.

In summary, the evolution equations of GRMHD can be written
in the following general form
\begin{equation}
{1 \over \sqrt{-g}} {\partial \sqrt{\gamma} \mathbf{U} \over \partial t} = - {1 \over
\sqrt{-g}} {\partial \sqrt{-g} \mathbf{F} \over \partial x^{i}} + \mathbf{S},
\label{eq2-41}
\end{equation}
where the quantities $\mathbf{U}$, $\mathbf{F}$, and $\mathbf{S}$ are
\begin{equation}
\mathbf{U} = \left[ \begin{array}{c}
D \\
S_{i} \\
\tau \\
B^{j}
\end{array} \right], \label{eq2-42}
\end{equation}

\begin{equation}
\mathbf{F} = \left[ \begin{array}{c}
D \tilde{v}^{i} \\
T^{j}_{i} \\
\alpha T^{ti} - D \tilde{v}^{i} \\
\tilde{v}^{i} B^{j} - \tilde{v}^{j} B^{i}
\end{array} \right], \label{eq2-43}
\end{equation}
and
\begin{equation}
\mathbf{S} = \left[ \begin{array}{c}
0 \\
T^{\mu \nu} \left( {\partial g_{\nu i} \over \partial x^{\mu}} -
\Gamma^{\sigma}_{\nu \mu} g_{\sigma i} \right) \\
\alpha \left( T^{\mu t} {\partial \ln \alpha \over \partial x^{\mu}}
- T^{\mu \nu} \Gamma^{t}_{\nu \mu} \right) \\
0^{j}
\end{array} \right], \label{eq2-44}
\end{equation}
where $0^{j} \equiv (0, 0, 0)^{T}$.

\section{Numerical Scheme}

Numerical evolution of the GRMHD equations involves determining the
fundamental MHD variables $\mathbf{P} = (\rho, p, v^{i}, B^{i})$,
called the ``primitive'' variables, at future times, given initial
values of $\mathbf{P}_{0}$. The evolution equations of GRMHD are
written in conserved form (e.g., Eqs. (\ref{eq2-41})-(\ref{eq2-44})).
They give the time derivatives for ``conserved'' variables
$\mathbf{U}(\mathbf{P})=(D, S_{i}, \tau, B^{i})$ in terms of source
variables $\mathbf{S}(\mathbf{P})$ and the divergence of flux
variables $\mathbf{F}$;
\begin{equation}
{\partial \mathbf{U} \over \partial t}  = - {\partial \mathbf{F} \over \partial x^{i}}
+ \mathbf{S} \equiv L(\mathbf{U}). \label{eq3-1}
\end{equation}

There are several ways to evolve the GRMHD equations numerically.
Conservative schemes evolve $\mathbf{U}$ with equation (\ref{eq3-1})
at each time step. The advantage of these schemes is
that highly accurate shock-capturing schemes can be applied to the
GRMHD equations. The disadvantage is that these schemes must recover
$\mathbf{P}$ by numerically solving the system of equations,
$\mathbf{U} = \mathbf{U}(\mathbf{P})$, after each time step. This
can be complicated. Conservative schemes for GRMHD have been
developed by Koide et al. (1999), Koide (2003),
 Gammie et al. (2003), Komissarov (2004), Duez et al.(2005),
Shibata \& Sekiguchi (2005) and Ant\'{o}n et al.(2006).

On the other hand non-conservative schemes such as ZEUS (Stone \&
Norman 1992) evolve variables which are more simple with respect to
$\mathbf{P}$, but whose evolving equations are not the same form of
Eq. (\ref{eq3-1}). In such schemes, high resolution shock-capturing
schemes cannot be applied and artificial viscosity must be used for
handling discontinuities. The advantage of these schemes is they
solve the internal energy equation instead of the energy equations
(Eq. (\ref{eq2-38})). This is an advantage in regions where the internal
energy is small compared to the total energy (such as highly
supersonic flows). Moreover, the recovery of $\mathbf{P}$ is fairly
straightforward. Non-conservative schemes for GRMHD have been
developed by De Villiers \& Hawley (2003) and Anninos et al. (2005).

Our GRMHD code described in detail in the following section employs
conservative schemes to solve the three-dimensional GRMHD equations
on uniform and non-uniform grids in each spatial direction (method of lines).
To maintain flexibility our GRMHD code is programmed to allow for
different boundary conditions, different coordinates (Cartesian,
Cylindrical and Spherical in RMHD and Boyer-Lindquist coordinates in
both non-rotating and rotating black holes), different spatial
reconstruction algorithms, different time advance algorithms, and
different recovery schemes.

\subsection{The reconstruction step}

In order to improve the spatial accuracy of the code, we
interpolate the primitive variables within computational
zones. These reconstructed variables are used to compute the
fluxes $\mathbf{F}$. For simplicity, we will consider the
one-dimensional case. The generalization to three dimensions
is straightforward. The primitive variables to the left and
right of the grid cell interface are $\mathbf{P}_{L}=
\mathbf{P}_{i+1/2-\epsilon}$ and $\mathbf{P}_{R}=\mathbf{P}_{i+1/2+\epsilon}$
respectively. We have implemented several reconstruction
schemes for computing $\mathbf{P}_{L}$ and $\mathbf{P}_{R}$.

\begin{description}
    \item[1)] Piecewise linear method (PLM) reconstruction

We use the minmod slope-limited linear interpolation method and
Monotonized central (MC) slope-limited linear interpolation method
(van Leer 1977). These methods give results with the second-order
accuracy at smooth regions and switch to first-order at local
extrema. For given primitive variables $a$, values of $a$ at the
left and right of the grid cell interface $a_{L}$ and $a_{R}$ are
computed according to
\begin{equation}
 a_{L} = a_{i} + \nabla a_{i} \Delta x/2
\end{equation}
\begin{equation}
 a_{R} = a_{i+1} - \nabla a_{i+1} \Delta x/2
\end{equation}
Here, $\nabla a$ is the slope-limited gradient of $a$.

In the minmod slope limited linear interpolation method,
\begin{equation}
\nabla a = \Delta x^{-1} minmod(a_{i+1}-a_{i}, a_{i}-a_{i-1}),
\end{equation}
\begin{equation}
minmod(a,b) = \left\{ \begin{array}{ll}
0 & \mbox{if $ab \le 0$}, \\
sign(a)min(|a|,|b|) & \mbox{otherwise}.
\end{array} \right.
\end{equation}

In the MC slope-limited linear interpolation method,

\begin{equation}
\nabla a = \Delta x^{-1} MC(a_{i+1}-a_{i}, a_{i}-a_{i-1}),
\end{equation}
\begin{equation}
MC(a,b) = \left\{ \begin{array}{ll}
0 & \mbox{if $ab \le 0$}, \\
sign(a)min(2 |a|, 2 |b|, |a+b|/2) & \mbox{otherwise}.
\end{array} \right.
\end{equation}

\item[2)] Convex, essentially non-oscillatory (CENO) reconstruction

 In this scheme, polynomial (quadratic) interpolation is used to
obtain the primitive variables of the left and right of the grid
cell interface. In smooth regions, these values are accurate to
third order in $\Delta x$. The scheme becomes first order at local
extrema. The details of this scheme are written in Del Zanna \&
Bucciantini (2002) and Del Zanna et al. (2003).

\item[3)] Piecewise parabolic method (PPM) reconstruction

In this scheme, the quadratic polynomial interpolation is used to
obtain the primitive variables to the left and right of the grid
cell interface. These reconstructed values are then modified such
that the parabolic profile defined by $a_{L}$, $a_{R}$ and $a_{i}$
is monotonic inside the grid cell. The modified interpolated values
at the grid cell interfaces define local Riemann problems. In the
regions near the contact discontinuities, the interpolation
procedure is  modified slightly to account sharp jumps. In the
vicinity of the local extrema, the scheme switches to a piecewise
constant approximation in order to avoid post shock oscillations. In
smooth regions, these values are accurate to fourth order in $\Delta
x$. The scheme becomes first order at local extrema. The details of
this scheme are written in Colella \& Woodward (1984) and the
relativistic version of the PPM algorithm is written in Mart\'{i}
and M\"{u}ller (1996).

\end{description}

\subsection{The Riemann solver step}

To calculate the numerical flux, we use the HLL (Harten, Lax, and
van Leer) approximate Riemann solver (Harten et al. 1983). 
The HLL approximate Riemann solver is one of the simplest
shock-capturing schemes because it does not require the eigenvector
of the characteristic matrix. However, when it couples with a
high-order reconstruction scheme, it has been shown to perform with
an accuracy comparable to more sophisticated solvers in shock tube
problems (Del Zanna \& Buccianitni 2002; Lucas-Serrano et al. 2004).
To calculate the HLL fluxes, one only needs to know a maximum
left-going wave speed $c_{+}$ and a maximum right-going wave speed
$c_{-}$ at the both sides of the grid cell interface.

From $\mathbf{P}_{R}$ and $\mathbf{P}_{L}$, we calculate
the maximum left-going wave speeds $c_{\pm, L}$, the maximum
right-going wave speeds $c_{\pm, R}$, the fluxes
$\mathbf{F}_{R} = \mathbf{F}(\mathbf{P}_{R})$ and
$\mathbf{F}_{L} = \mathbf{F}(\mathbf{P}_{L})$,
and conserved variables $\mathbf{U}_{R} = \mathbf{U}
(\mathbf{P}_{R}) $ and $\mathbf{U}_{L} = \mathbf{U}
(\mathbf{P}_{L})$. Defining $c_{\mathrm{max}} \equiv
max(0, c_{+,R}, c_{+,L})$ and $c_{\mathrm{min}} \equiv
- min(0, c_{-,R}, c_{-,L})$, the HLL flux is given by
\begin{equation}
\mathbf{F}_{i+1/2} = {c_{\mathrm{min}} \mathbf{F}_{R} + c_{\mathrm{max}}
\mathbf{F}_{L} - c_{\mathrm{min}}c_{\mathrm{max}} (\mathbf{U}_{R} - \mathbf{U}_{L})
\over c_{\mathrm{max}} + c_{\mathrm{min}}} .
\end{equation}

We calculate the wave speeds $c_{\pm}$ by the same method as Gammie et al. 
(2003) and Duez et al. (2005). The HLL approximate Riemann
solver requires only the maximum wave speeds in either direction
along the three coordinate axes. To determine the wave speeds in the
$x^{1}$ direction, one solves the dispersion relation for MHD waves
with wave vectors of the form
\begin{equation}
k_{\mu} = (- \omega, k_{1}, 0, 0).
\end{equation}
The wave speed is the phase speed $\omega / k_{1}$. The speeds along
the $x^{2}$ and $x^{3}$ direction are calculated in a similar way.
The full dispersion relation for fast and slow modes is a
fourth-order polynomial. It is difficult to solve the full
dispersion relation. Gammie et al. (2003) 
has proposed replacing
 the full dispersion relation by the simpler approximate
expression which overestimates the maximum speeds by a factor of
$\le 2$ (it makes the evolution more diffusive but more stable). In
the frame comoving with the fluid, the approximate dispersion relation
for slow and fast modes is written by
\begin{equation}
\omega^{2} = [v_{A}^{2}+ c_{s}^{2} (1 - v_{A}^{2})] k^{2},
\end{equation}
where $c_{s} = \sqrt{\Gamma p/ (\rho h)}$ is the sound speed,
$v_{A}$ is the Alfv\'{e}n
speed, $v^{2}_{A} = b^{2}/ (\rho h + b^{2})$.

\subsection{Constrained transport}

High resolution shock-capturing scheme can successfully solve
many problems involving various kinds of discontinuities. However
these schemes do not guarantee $\mathbf{\nabla} \cdot \mathbf{B} =0$
in multidimensional simulations. Therefore we need to use
constrained transport schemes to evolve the induction equation while
maintaining $\mathbf{\nabla} \cdot \mathbf{B} =0$.

Several approaches and schemes have been proposed to maintain
$\mathbf{\nabla} \cdot \mathbf{B} =0$ (e.g., T\'{o}th 2002). We use
the flux-interpolated, constrained transport (flux-CT) scheme
introduced by T\'{o}th (2002) and used by Gammie et al. (2003) 
and Duez et al. (2005). This scheme involves replacing the numerical
flux of the induction equation computed at each point with linear
combinations of the numerical fluxes computed at that point and
neighboring points. The merit of this scheme is that it is naturally
coupled with a Godunov-type scheme (T\'{o}th 2002; Gammie et al 2003).
It does not need the staggered variables which are needed in the
constrained transport scheme by Evans \& Hawley (1988), and the
advantage of its usage with higher order implementation of the
divergence free condition is discussed in Londrillo \& Del Zanna
(2004).

\subsection{Time advance step}

In the time advance step, we get the updated values of the
conserved variables at the next time levels ($\mathbf{U}^{n+1}$).
We use a multistep TVD Runge-Kutta (RK) method developed by
Shu \& Osher (1988) that can provide second (RK2) and third (RK3)
order accuracy in time. The explicit form of the algorithms is:

\begin{enumerate}
    \item Prediction step (common for RK2 and RK3):

\begin{equation}
\mathbf{U}^{(1)} = \mathbf{U}^{n} + {\Delta t} L(\mathbf{U}^{n}). \label{eq3-4-1}
\end{equation}

   \item Depending on the accuracy of the time advance scheme do:

RK2:

\begin{equation}
\mathbf{U}^{n+1} = {1 \over a} [ b \mathbf{U}^{n} + \mathbf{U}^{(1)} +
\Delta t L(\mathbf{U}^{(1)}) ], \label{eq3-4-2}
\end{equation}
in this case $a=2$ and $b=1$.

RK3:
\begin{equation}
\mathbf{U}^{(2)} = {1 \over a} [ b \mathbf{U}^{n} + \mathbf{U}^{(1)} +
\Delta t L(\mathbf{U}^{(1)}) ], \label{3-4-3}
\end{equation}
\begin{equation}
\mathbf{U}^{n+1} = {1 \over b} [ \mathbf{U}^{n} + 2 \mathbf{U}^{(2)} +
2 \Delta t L(\mathbf{U}^{(2)}) ], \label{eq3-4-4}
\end{equation}
in this case $a=4$ and $b=3$.
\end{enumerate}

\subsection{Recovery of primitive variables}

The conservative MHD scheme for GRMHD requires a method for
transforming between conserved variables $\mathbf{U}$ and primitive
variables $\mathbf{P}$. The forward transformation (from primitive
to conserved variables) has a closed-form solution, but the inverse
transformation (from conserved to primitive variables) requires the
solution of a set of five nonlinear equations. Having computed
$\mathbf{U}$ at the new timestep, we must calculate primitive
variables $\mathbf{P}$ at the new time.

For the recovery we prepare two methods, Koide's 2D method
(Koide et al. 1999) and Noble's 2D method (Noble et al. 2006).

\begin{enumerate}
    \item Koide's 2D method

In Koide et al. (1999) they solve two nonlinear, simultaneous
algebraic equations with two independent variables $x \equiv
\gamma-1$ and $y \equiv \gamma (\mathbf{v} \cdot \mathbf{B})$.
In these equations they use $\Gamma$-law EOS (see Eq. (\ref{eq2-40}))
\begin{eqnarray}
&&x(x+2) \left[ \Gamma R x^{2} + (2 \Gamma R -d )x + \Gamma R - d + e + {\Gamma
\over 2}
y^{2} \right]^{2} \cr
&& =(\Gamma x^{2} + 2 \Gamma x +1)^{2} [f^{2} (x+1)^{2}+ 2 \sigma y + 2
\sigma x y + g^{2}y^{2}], \label{eq3-5-1}
\end{eqnarray}
\begin{eqnarray}
&&\left[ \Gamma  ( R-g^{2} ) x^{2} + (2 \Gamma R- 2 \Gamma g^{2} -d)x +
\Gamma R-d +e-g^{2} + { \Gamma \over 2 } y^{2} \right] y \cr
&&= \sigma (x+1)(\Gamma x^{2} + 2 \Gamma x +1), \label{eq3-5-2}
\end{eqnarray}
where $R = D + \tau$, $d= (\Gamma-1)D$, $e=(1-\Gamma/2)B^{2}$, $f=\mathbf{S}$,
$g= \mathbf{B}$, and $\sigma = \mathbf{B} \cdot \mathbf{S}$.
Note that, in the absence of the magnetic field $B^{i}$,
Eq (\ref{eq3-5-1}) reduces to the well-known relativistic
hydrodynamic one derived by Duncan and Hughes (1994), and
Eq (\ref{eq3-5-2}) becomes a trivial equation. These algebraic
equations are solved at each grid cell using a 2-variable
Newton-Raphson iteration method. The primitive variables then
are calculated easily from $x$, $y$, $D$, $\mathbf{S}$,
$\tau$, and $\mathbf{B}$, using
\begin{equation}
\gamma = 1+x,
\end{equation}
\begin{equation}
p={(\Gamma -1 ) [\tau - x D - (2-1/\gamma^{2})B^{2}/2+(y/\gamma)^{2}/2] \over
[\gamma x (x+2) +1]}, \label{3-5-3}
\end{equation}
and
\begin{equation}
\mathbf{v}={\mathbf{S} + (y/\gamma)\mathbf{B} \over D + \{ \tau + p + B^{2}/2
\gamma^{2} + (y/ \gamma)^{2}/2 \} }. \label{3-5-4}
\end{equation}
This method is identical to that used in special relativistic MHD simulations
(Koide, Nishikawa, \& Mutel 1996; Koide 1997).

    \item Noble's 2D method

In Noble et al. (2006) they have tried six numerical methods for
performing the inverse transformation and discuss the mathematical
properties of them. We use one method which is recommended in
Noble et al. (2006). This method solves two simple algebraic equations
simultaneously for two independent variables $W \equiv \gamma^{2}h$
and $v^{2}$
\begin{equation}
S^{2} = (W + B^{2}) v^{2} - {(B^{2} + 2 W)
(\mathbf{S}\cdot \mathbf{B})^{2} \over W^{2}}, \label{3-5-5}
\end{equation}
and
\begin{equation}
\tau = {B^{2} \over 2}(1 + v^{2}) + {\mathbf{S} \cdot \mathbf{B} \over 2 W} + W - D
- p(u, \rho). \label{3-5-6}
\end{equation}
and
\end{enumerate}
Note that in Eq. (\ref{3-5-6}) we can use any EOS. If we adopt a
$\Gamma$-law EOS (Eq. (\ref{eq2-40})), Eq. (\ref{3-5-6}) can
be written as
\begin{equation}
\tau = {B^{2} \over 2}(1 + v^{2}) + {\mathbf{S} \cdot \mathbf{B} \over 2 W} + W - D -
\left( {\Gamma-1 \over \Gamma} [ (1-v^{2}) W - \rho ] \right).
\end{equation}
These algebraic equations are solved at each grid cell using a
2-variable Newton-Raphson iteration method. From $W$ and $v^{2}$
the primitive variables $\rho$ and $p$ (or $u$) are
calculated easily.

\section{Code Tests}

In this section, we test the capabilities of our new GRMHD code. The
tests are non-relativistic, special relativistic and general
relativistic.

\subsection{Linear Alfv\'{e}n wave propagation}

The first test considers the propagation of a small-amplitude
Alfv\'{e}n wave in one dimension of Cartesian coordinates. The
initial conditions are $\rho=1.0$, $p=1.0$, $v^{x}=0.0$, $v^{y}=A
\cos(k x)$, $v^{z}=0.0$, $B^{x}=B_{0}$, $B^{y}=-B_{0}(A/v_{A}) \cos
(k x)$, and $B^{z}=0.0$, where $k=2 \pi$ and $A$ is the amplitude.
We use the parameters $B_{0}=1.0$ and $A = 0.01 $. The fluid
satisfies a $\Gamma$-law EOS with $\Gamma = 4/3$. The computational
domain is $0 \le x \le 1.0$ and the boundary condition is periodic.

The simulation runs for a single wave period $2 \pi/ \omega$
($t_{end}=2 \pi/ \omega$),
so that a perfect scheme would return to its initial state.
We measure the $\mathit{L}_{1}$ norm of the difference between
the final state and the initial state for each primitive
variables $Q$ such as
\begin{equation}
L_{1}(\delta Q) = \int_{i=1}^{N} \left| Q (t=0) - Q \left( t= t_{end}
\right) \right| dx \label{eq4-1}
\end{equation}
as a function of the computational zone number $N$.

\begin{figure}[ht]
\epsscale{0.8} \plotone{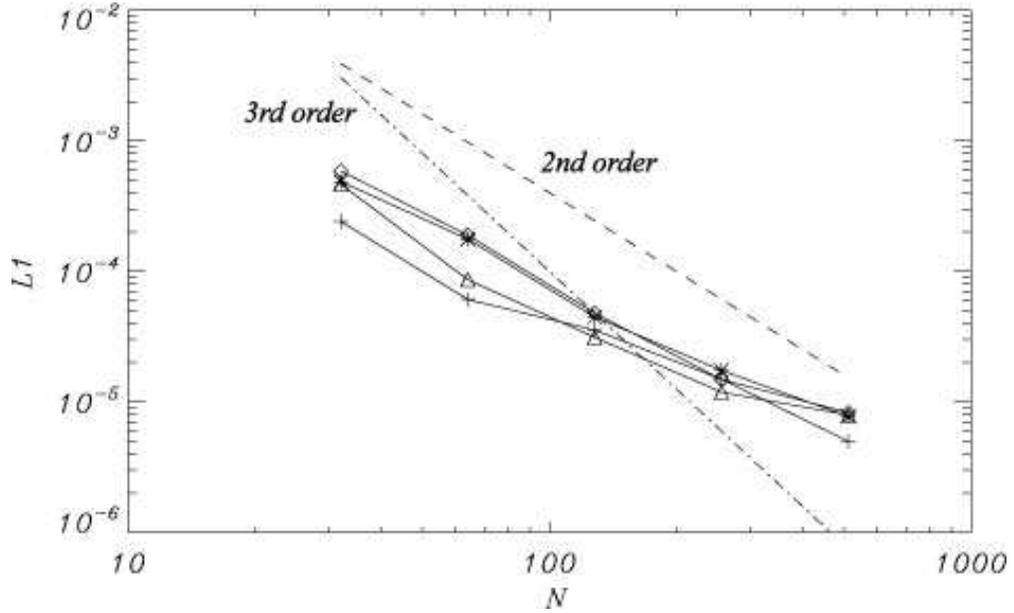} \caption{ $L_{1}$ norm of the error
in $v^{y}$ for a linear Alfv\'{e}n wave propagation as a function of
computational zone number, $N$, for the MC slope limiter (plus), the minmod slope
limiter (asterisk), convex CENO (open diamond) and PPM
reconstructions (open triangle). The straight lines show the slope
expected for second-order convergence (dashed line) and third-order
convergence (dash-dotted line). \label{f1}}
\end{figure}

Figure \ref{f1} shows the $L_{1}$ norm of the error in $v^{y}$
($Q=v^{y}$ in Eq (\ref{eq4-1})) for runs using the MC slope limiter,
minmod slope limiter, CENO, and PPM reconstructions as the
computational zone number $N$ is increased. Courant number is 0.5 in
all simulations. All reconstruction schemes show that the global
order of convergence of the code is second order in small
computational zone number $N$ and tend to flatter than second order
convergence. Even we use the higher order reconstruction schemes
such as CENO or PPM reconstructions, the global order of convergence
of the code is second order. It is because we use the flux-CT scheme
to maintain divergence-free magnetic field. It has second-order
accuracy. Even all reconstruction schemes show the second order
convergence, MC slope-limiter and PPM reconstructions are more
accurate than minmod slope limiter and CENO reconstruction schemes.

\subsection{Relativistic MHD shock-tube tests}

Shock-tube tests are the most basic test problems for MHD (HD)
codes. Large sets of test-problems in relativistic MHD have been
performed over the years (i.e., Komissarov 1999a; Balsara 2001). In
some test problems the exact solution of the Riemann problem in
relativistic MHD has been calculated by Giacomazzo \& Rezzolla
(2006).

We perform eight simulations of $B^{x} \neq 0$ cases which exact solutions
obtained by Giacomazzo \& Rezzolla (2006).
Therefore we can compare the simulation results with the exact
solutions directly. All tests start with discontinuous initial data
at $x=0$ (see Table \ref{table1}) and with homogenous profiles on
either side in Cartesian coordinates. We simulate from $t=0$ to
$t=t_{\rm final}$ with  different reconstruction schemes, where
$t_{\rm final}$ is specified in Table \ref{table1} for each case.
The fluid satisfies a $\Gamma$-law EOS. In all cases we use 400 computational
zones with a Courant factor of $0.5$.

\begin{figure}[ht]
\epsscale{1.0} \plotone{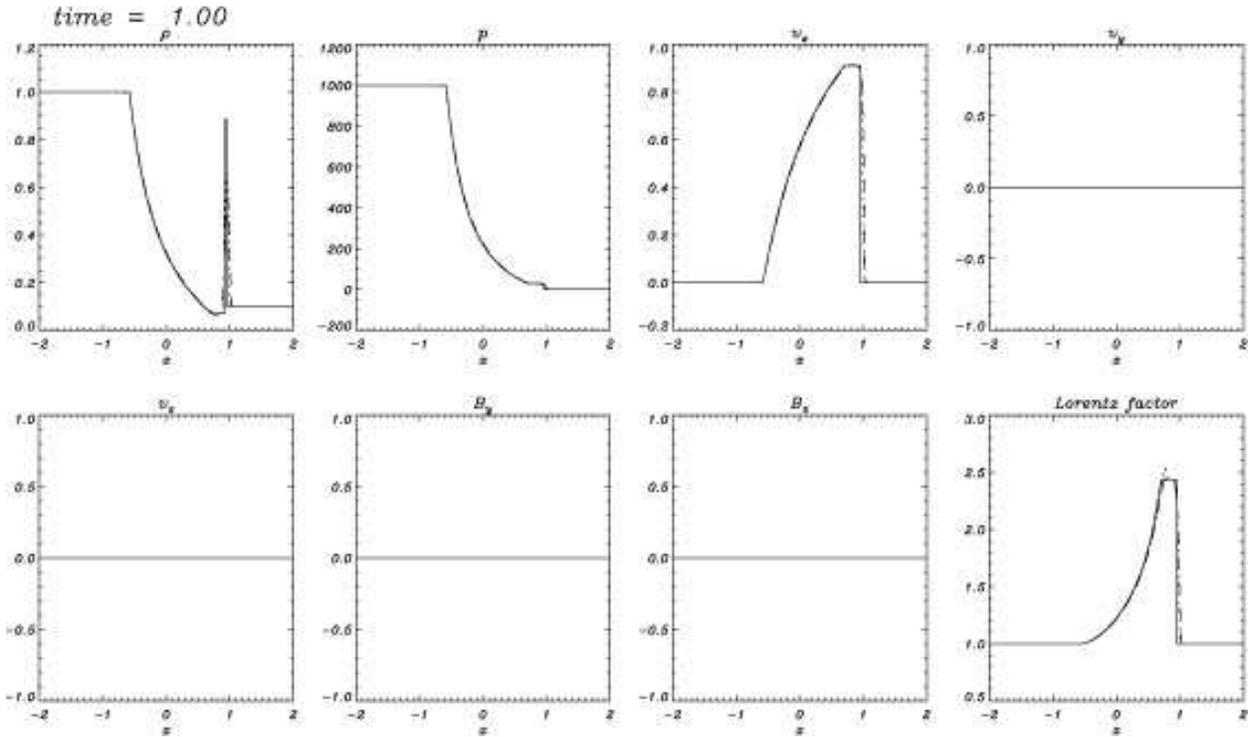} \caption{Simulation results of
Komissarov shock-tube test 1 at time $t=1.0$ using the MC slope
limiter (dotted lines) and the minmod slope limiter (dashed lines)
reconstructions. The solid lines are the exact solution. The results
are composed of a left-going fast rarefaction,  a contact
discontinuity, and  a right-going fast shock. \label{f3a}}
\end{figure}

\begin{figure}[ht]
\plotone{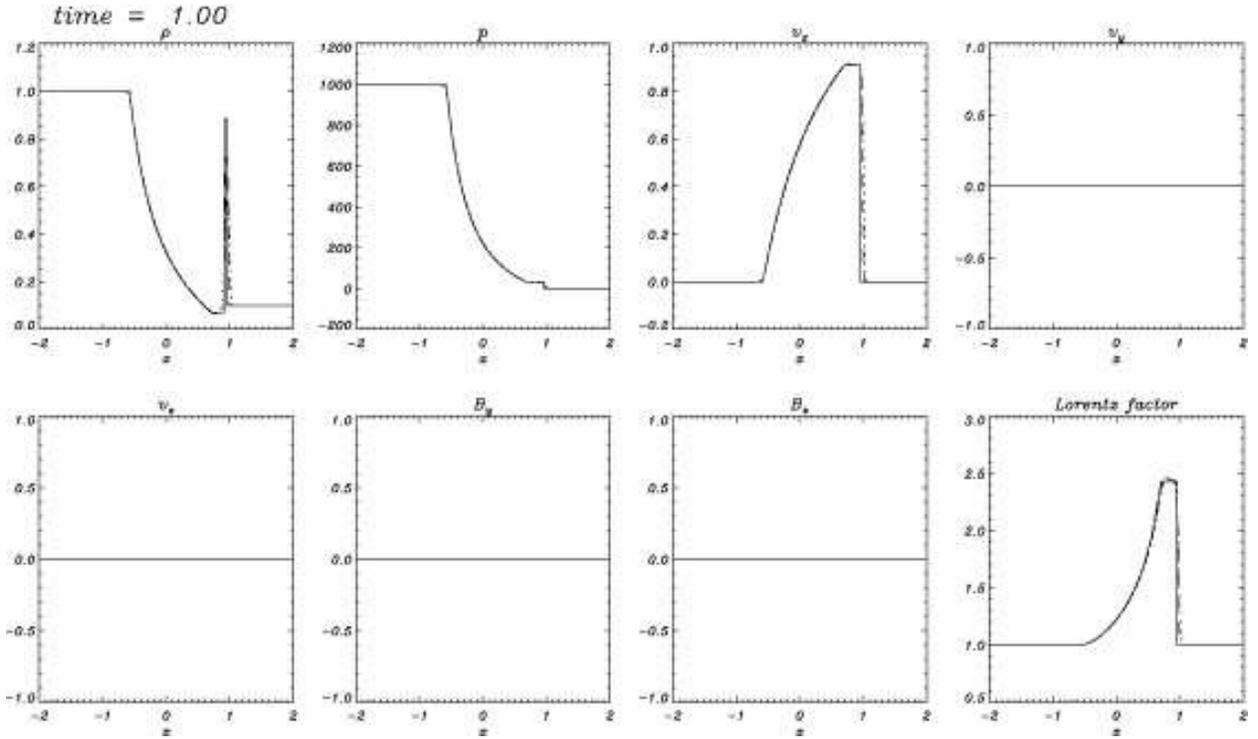} \caption{ Simulation results of Komissarov
shock-tube test 1 at time $t=1.0$ using the CENO (dotted lines) and
the PPM (dashed lines) reconstructions. The solid lines are the
exact solution. The results are composed of a left-going fast
rarefaction,  a contact discontinuity, and  a right-going fast
shock. \label{f3b}}
\end{figure}

\begin{figure}[ht]
\plotone{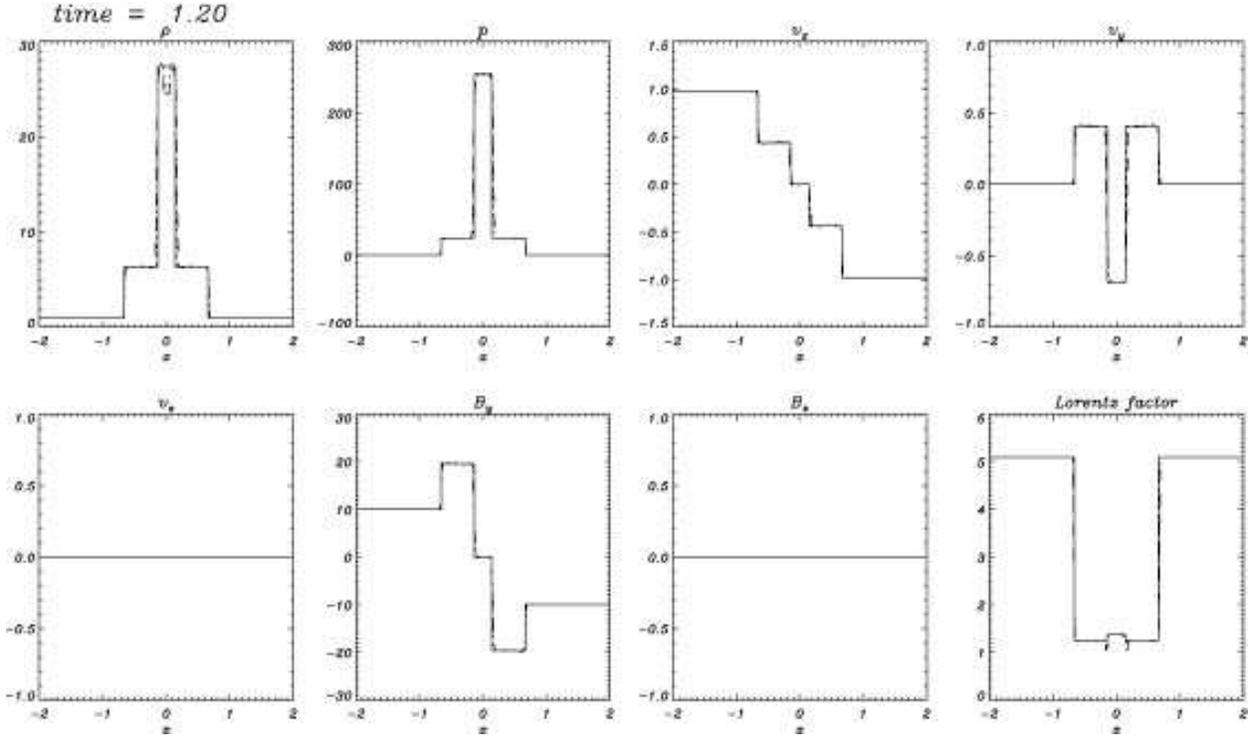} \caption{ Simulation results of Komissarov
collision test at time $t=1.5$ using the MC slope limiter (dotted
lines) and the minmod slope limiter (dashed lines) reconstructions.
The solid lines are the exact solution. The results are composed of
a left-going fast shock, of a left-going slow shock,  a right-going
slow shock, and  a right-going fast shock. \label{f4a}}
\end{figure}

\begin{figure}[ht]
\plotone{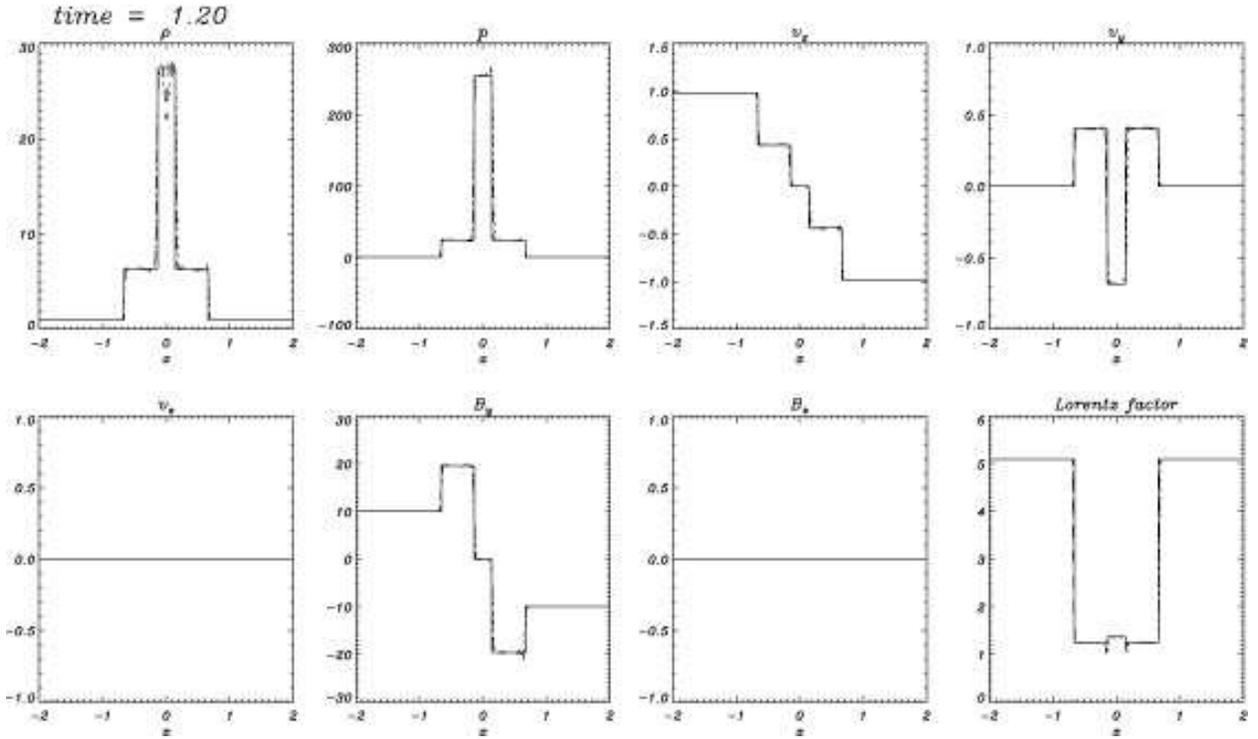} \caption{ Simulation results of Komissarov
collision test at time $t=1.5$ using the CENO (dotted lines) and the
PPM (dashed lines) reconstructions. The solid lines are the exact
solution. The results are composed of a left-going fast shock, of a
left-going slow shock, a right-going slow shock, and a right-going
fast shock. \label{f4b}}
\end{figure}

\begin{figure}[ht]
\plotone{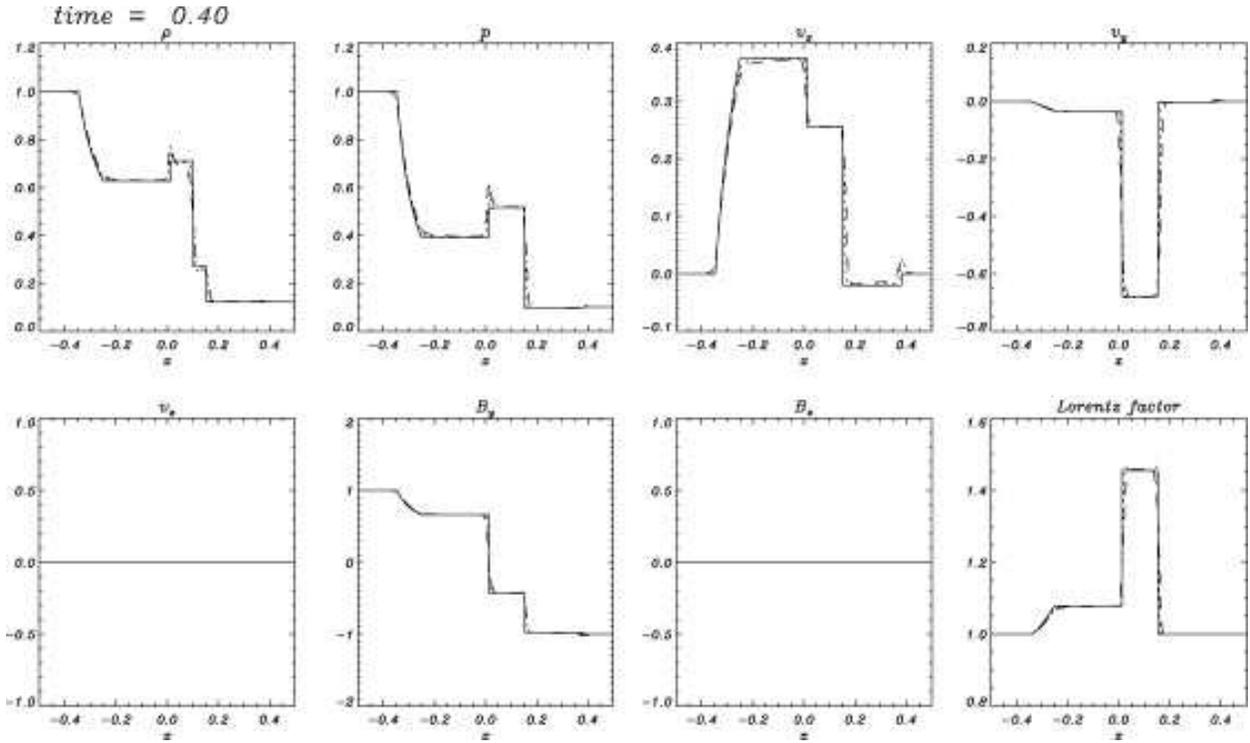}
\caption{ Simulation results of Balsara test 1 at time $t=0.4$
using the MC slope limiter (dotted lines) and the minmod slope
limiter (dashed lines) reconstructions. The solid lines are the
exact solutions. The results are composed of a left-going fast
rarefaction, of a left-going slow shock, of a contact
discontinuity, of a right-going slow shock and of a right-going
fast rarefaction. \label{f5a}}
\end{figure}

\begin{figure}[ht]
\plotone{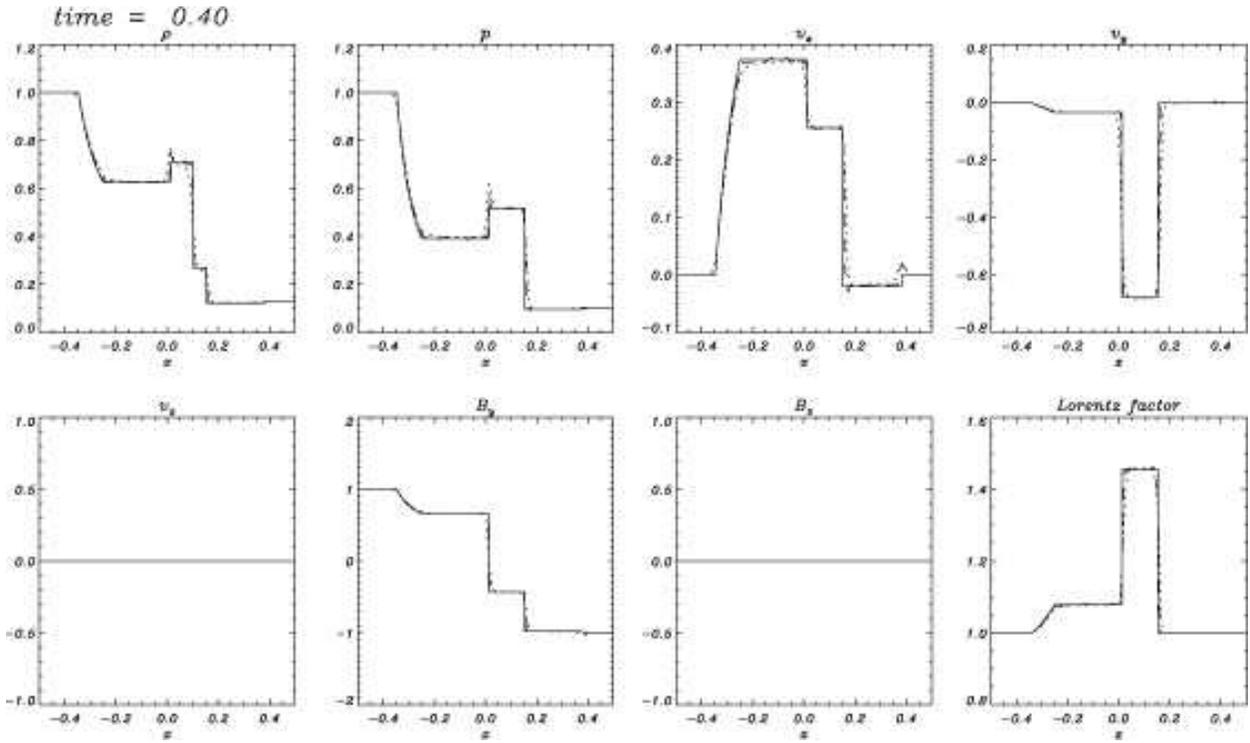} \caption{ Simulation results of Balsara test 1 at
time $t=0.4$ using the CENO (dotted lines) and the PPM (dashed
lines) reconstructions. The solid lines are the exact solutions. The
results are composed of a left-going fast rarefaction, of a
left-going slow shock, a contact discontinuity, of a right-going
slow shock, and  a right-going fast rarefaction. \label{f5b}}
\end{figure}

\begin{figure}[ht]
\plotone{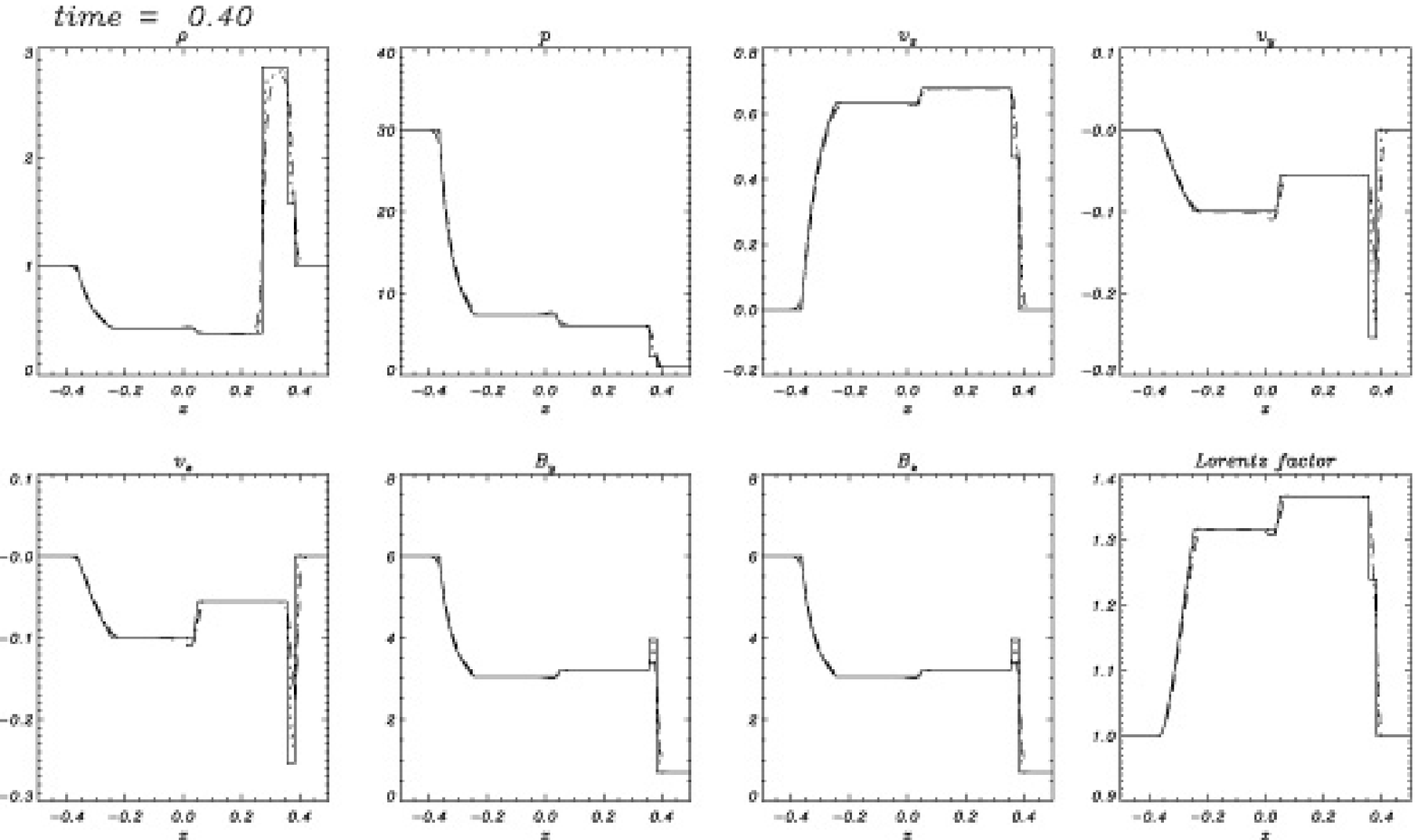} \caption{ Simulation results of Balsara test 2 at
time $t=0.4$ using the MC slope limiter (dotted lines) and the
minmod slope limiter (dashed lines) reconstructions. The solid lines
are the exact solutions. The results are composed of two left-going
fast and slow rarefactions, a contact discontinuity, and two
right-going fast and slow shocks. \label{f6a}}
\end{figure}

\begin{figure}[ht]
\plotone{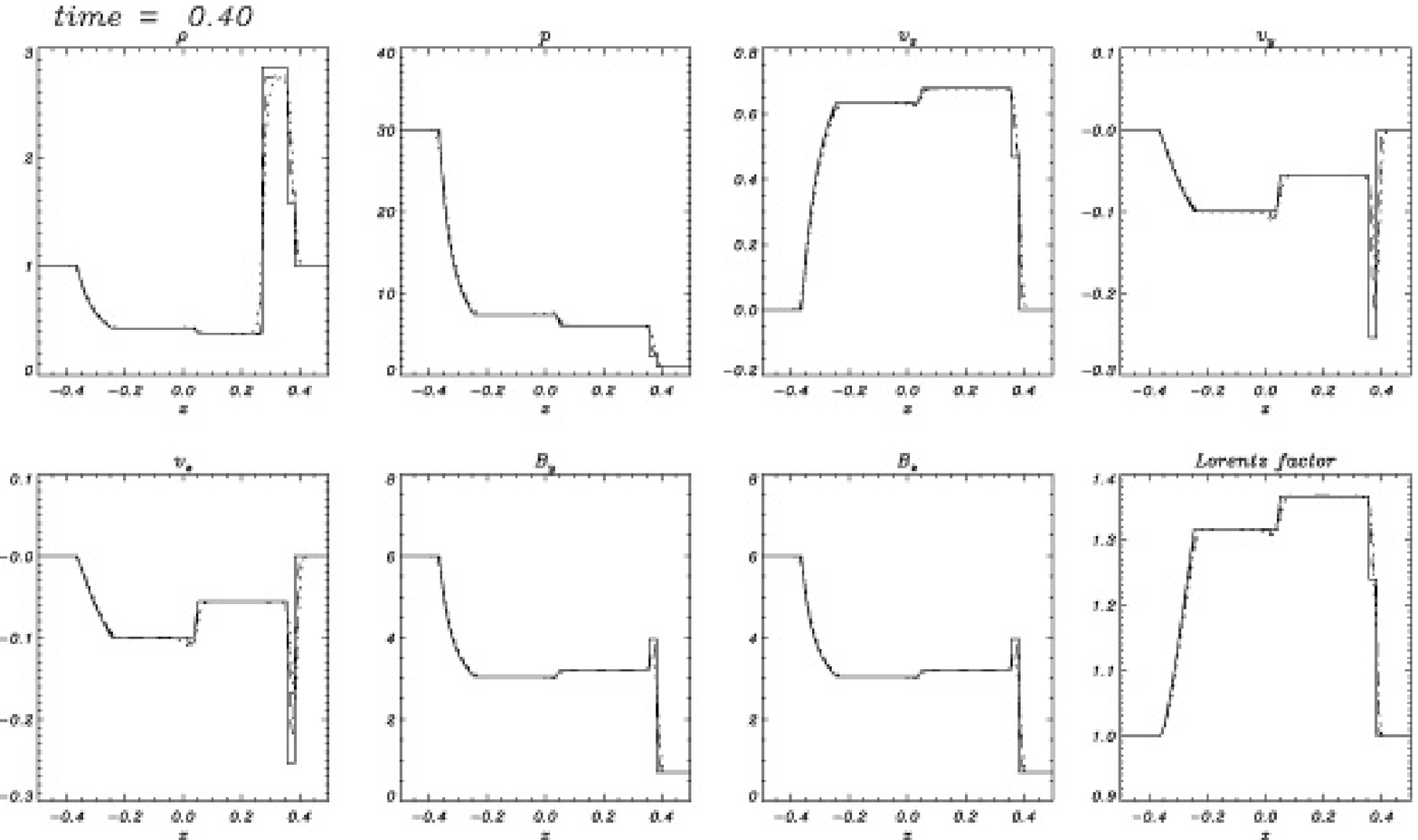} \caption{ Simulation results of Balsara test 2 at
time $t=0.4$ using the CENO (dotted lines) and the PPM (dashed
lines) reconstructions. The solid lines are the exact solutions. The
results are composed of two left-going fast and slow rarefactions, a
contact discontinuity, and two right-going fast and slow shocks.
\label{f6b}}
\end{figure}

\begin{figure}[ht]
\plotone{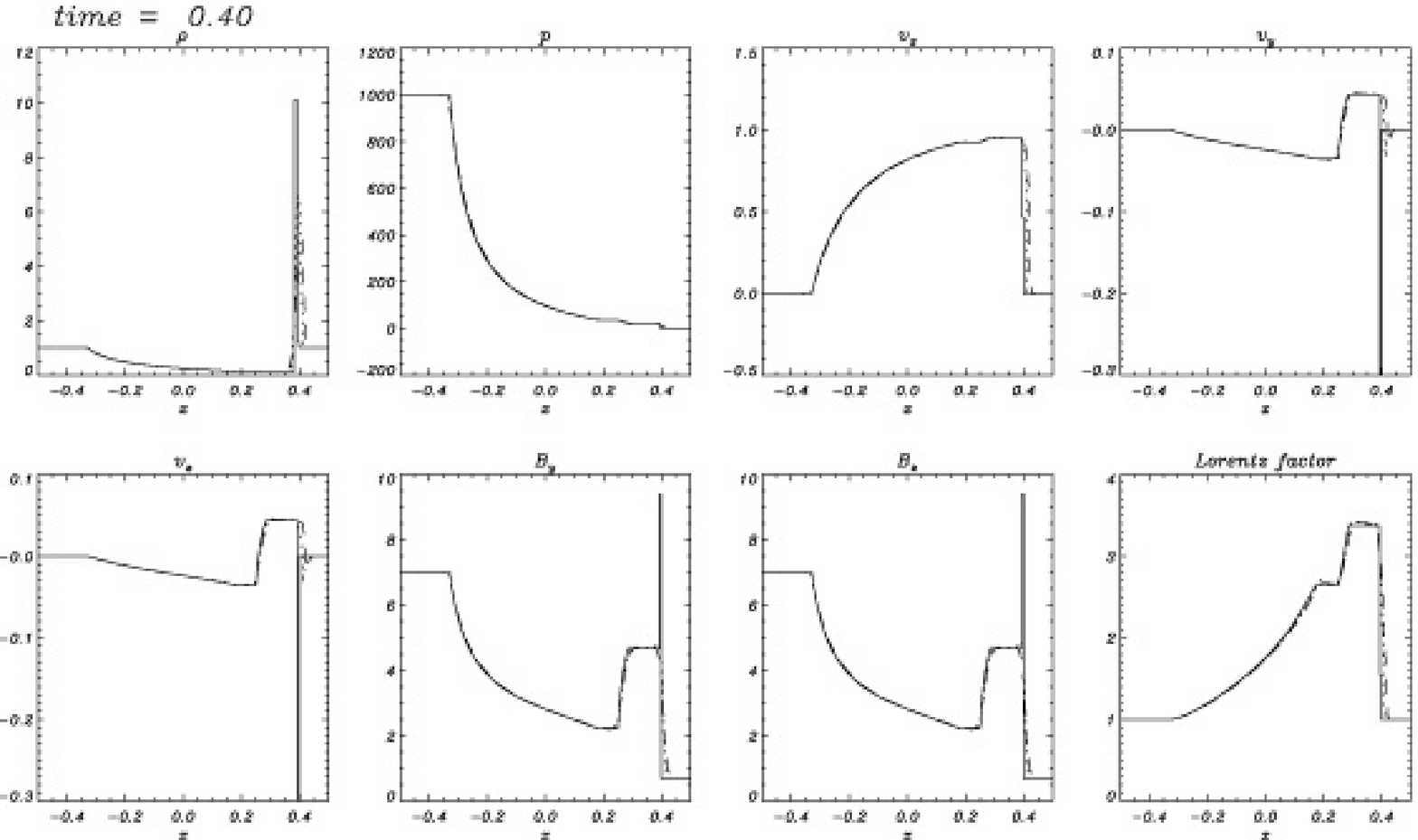} \caption{ Simulation results of Balsara test 3 at
time $t=0.4$ using the MC slope limiter (dotted lines) and the
minmod slope limiter (dashed lines) reconstructions. The solid lines
are the exact solution. The results are composed of two left-going
fast and slow rarefactions, a contact discontinuity, and  two
right-going fast and slow shocks. \label{f7a}}
\end{figure}

\begin{figure}[ht]
\plotone{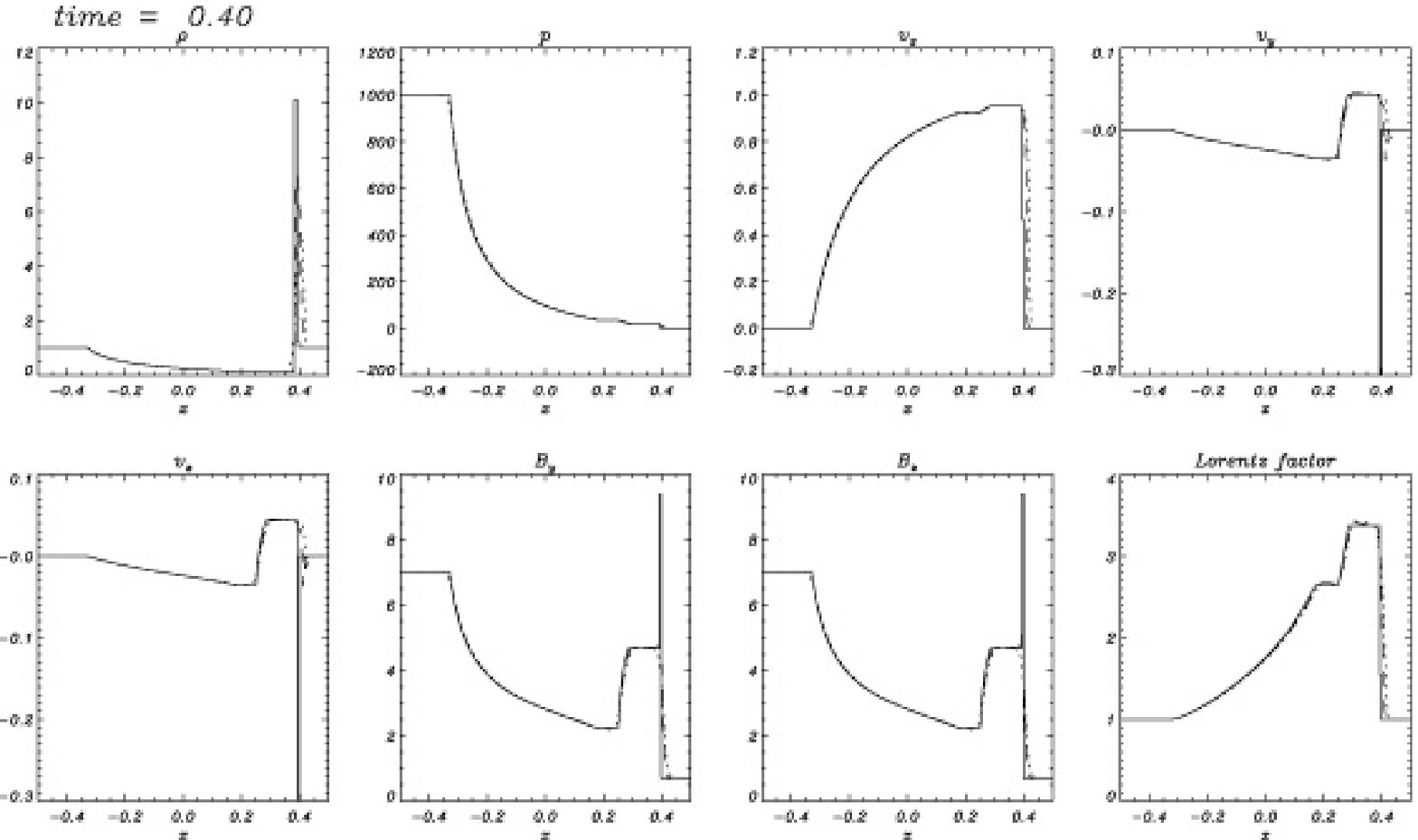} \caption{ Simulation results of Balsara test 3 at
time $t=0.4$ using the CENO (dotted lines) and the PPM (dashed
lines) reconstructions. The solid lines are the exact solution. The
results are composed of two left-going fast and slow rarefactions, a
contact discontinuity, and two right-going fast and slow shocks.
\label{f7b}}
\end{figure}

\begin{figure}[ht]
\plotone{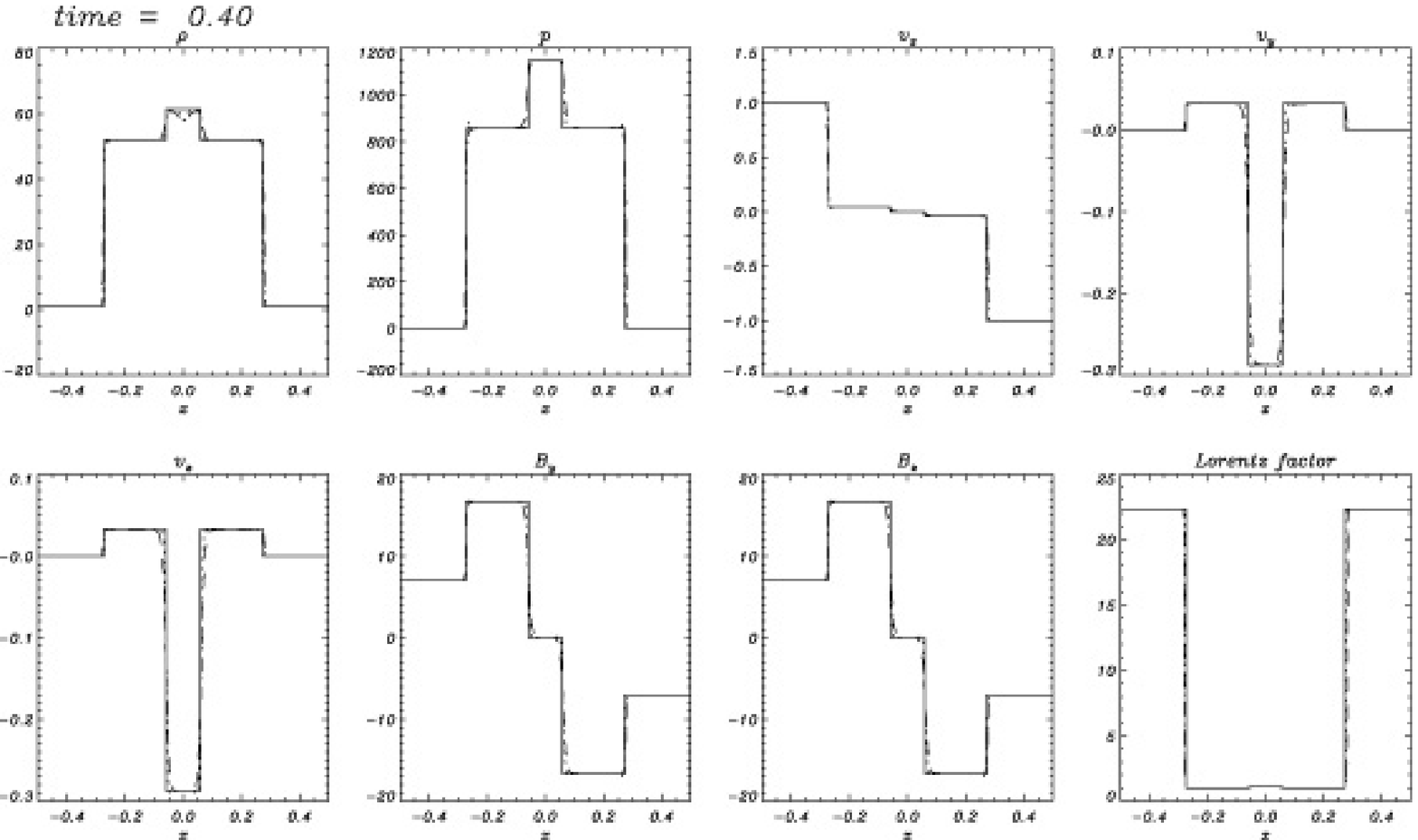} \caption{ Simulation results of Balsara test 4 at
time $t=0.4$ using the MC slope limiter (dotted lines) and the
minmod slope limiter (dashed lines) reconstructions. The solid lines
are the exact solution. The results are composed of two left-going
fast and slow shocks, and two right-going fast and slow shocks.
\label{f8a}}
\end{figure}

\begin{figure}[ht]
\plotone{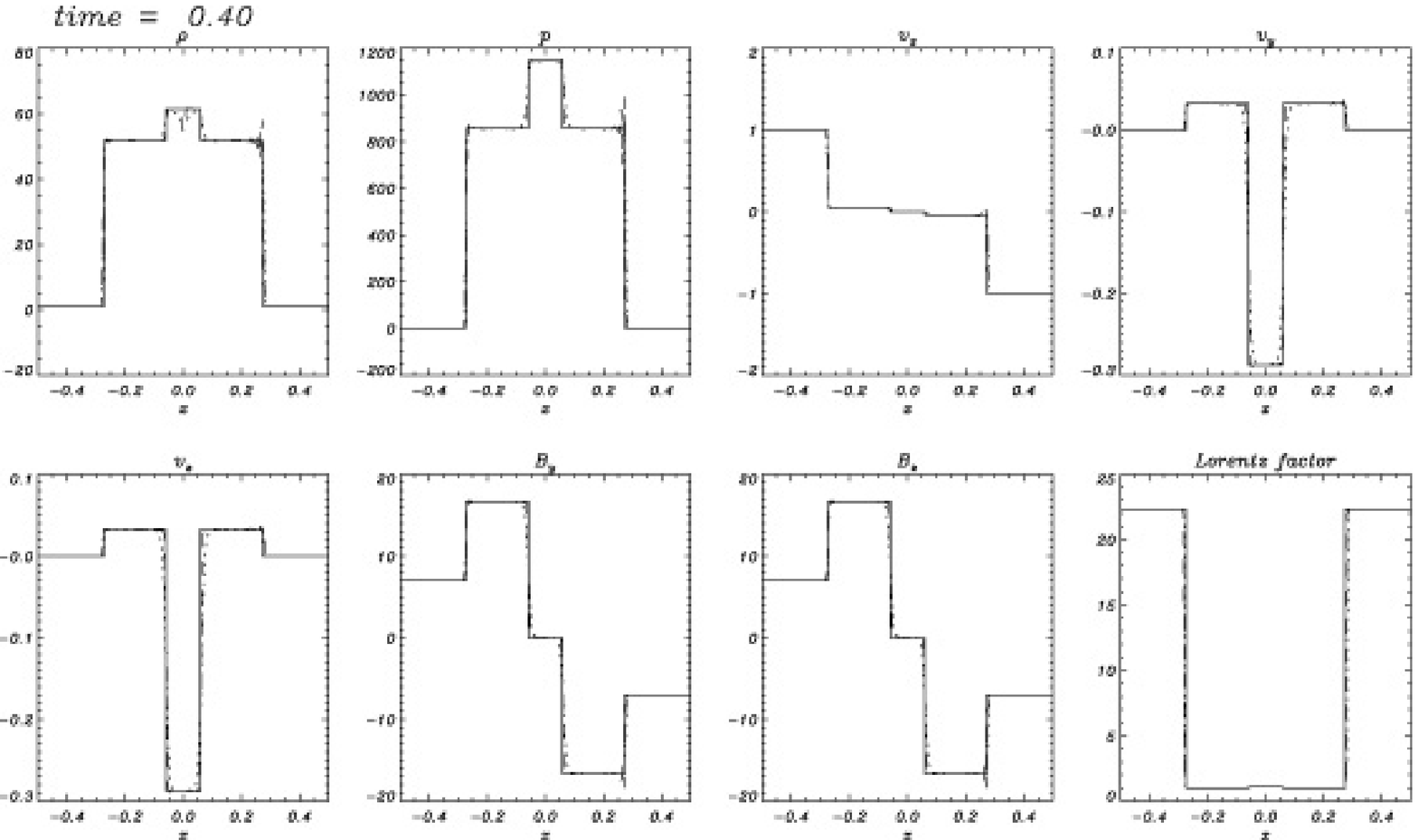} \caption{ Simulation results of Balsara test 4 at
time $t=0.4$ using the CENO (dotted lines) and the PPM (dashed
lines) reconstructions. The solid lines are the exact solution. The
results are composed of two left-going fast and slow shocks, and
two right-going fast and slow shocks. \label{f8b}}
\end{figure}

\begin{figure}[ht]
\plotone{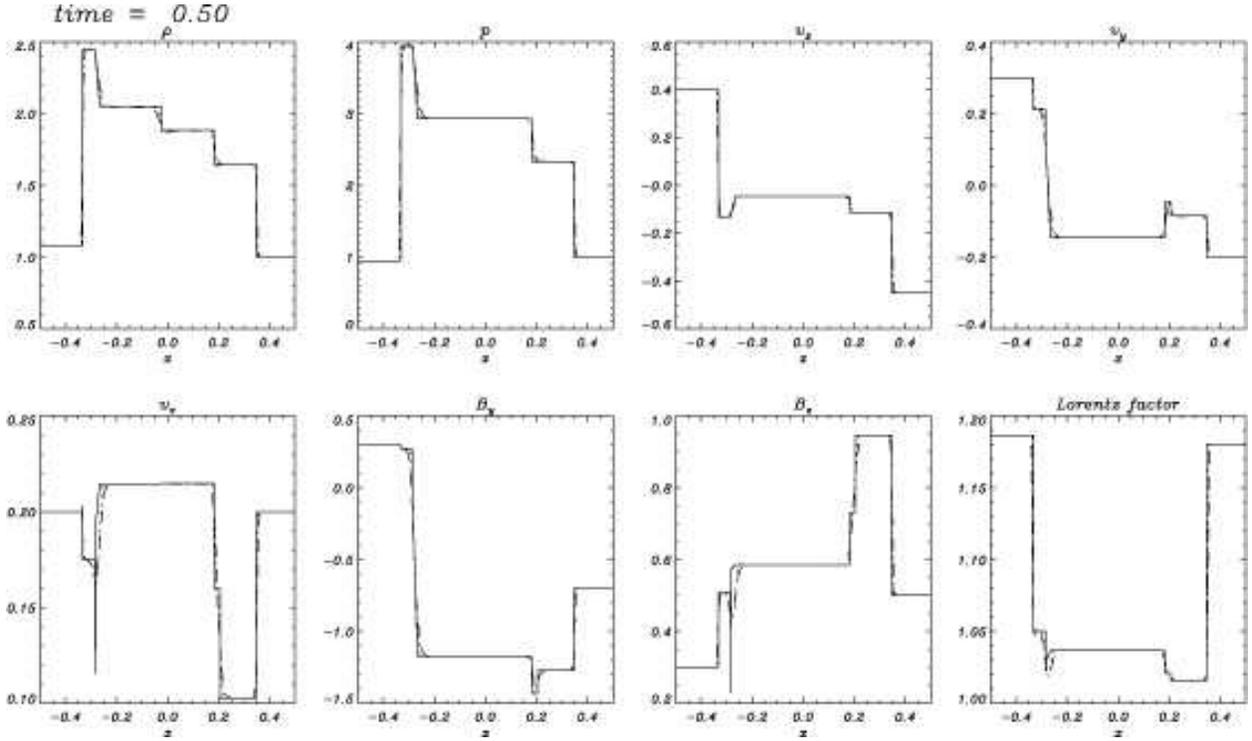} \caption{ Simulation results of Balsara test 5 at
time $t=0.5$ using the MC slope limiter (dotted lines) and the
minmod slope limiter (dashed lines) reconstructions. The solid lines
are the exact solution. The results are composed of a left-going
fast shock,  a left-going Alfv\'{e}n discontinuity, a left-going
slow rarefaction, a contact discontinuity, a right-going slow shock,
a right-going Alfv\'{e}n discontinuity, and  a right-going fast
shock. \label{f9a}}
\end{figure}

\begin{figure}[ht]
\plotone{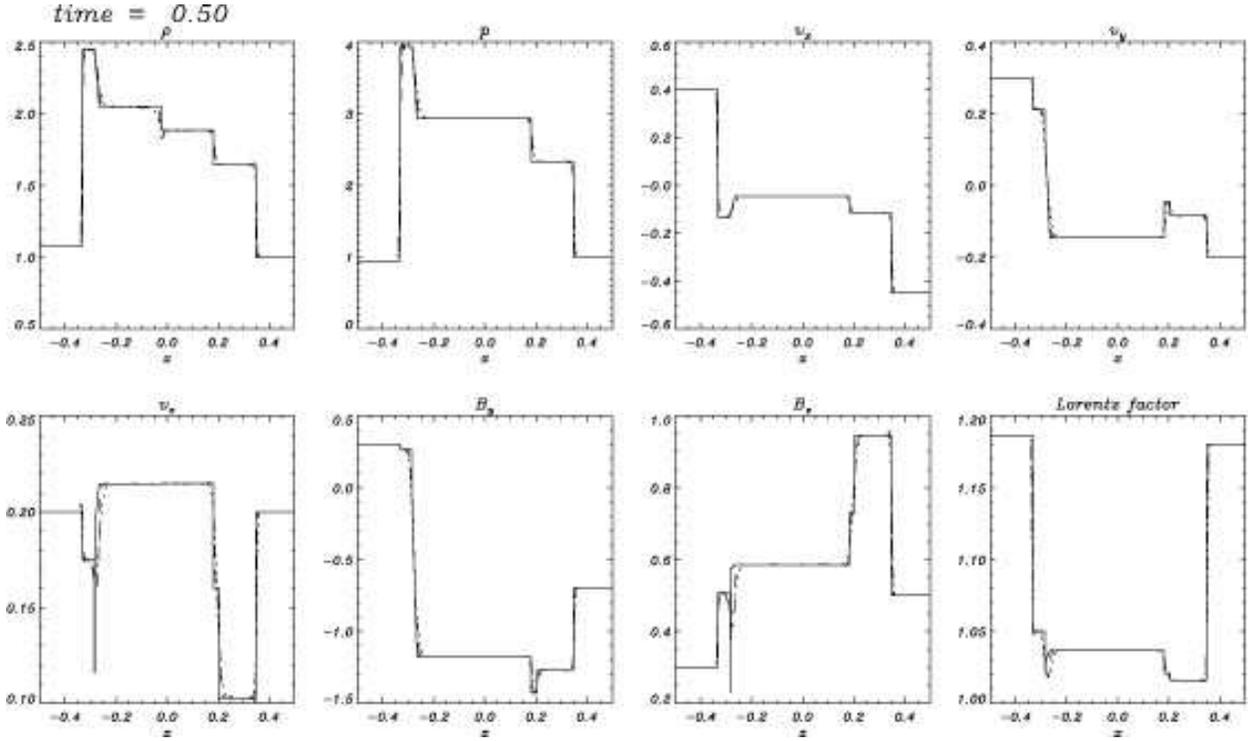} \caption{ Simulation results of Balsara test 5 at
time $t=0.5$ using the CENO (dotted lines) and the PPM (dashed
lines) reconstructions. The solid lines are the exact solution. The
results are composed of a left-going fast shock, a left-going
Alfv\'{e}n discontinuity, a left-going slow rarefaction, a contact
discontinuity, a right-going slow shock, a right-going Alfv\'{e}n
discontinuity, and  a right-going fast shock. \label{f9b}}
\end{figure}

\begin{figure}[ht]
\plotone{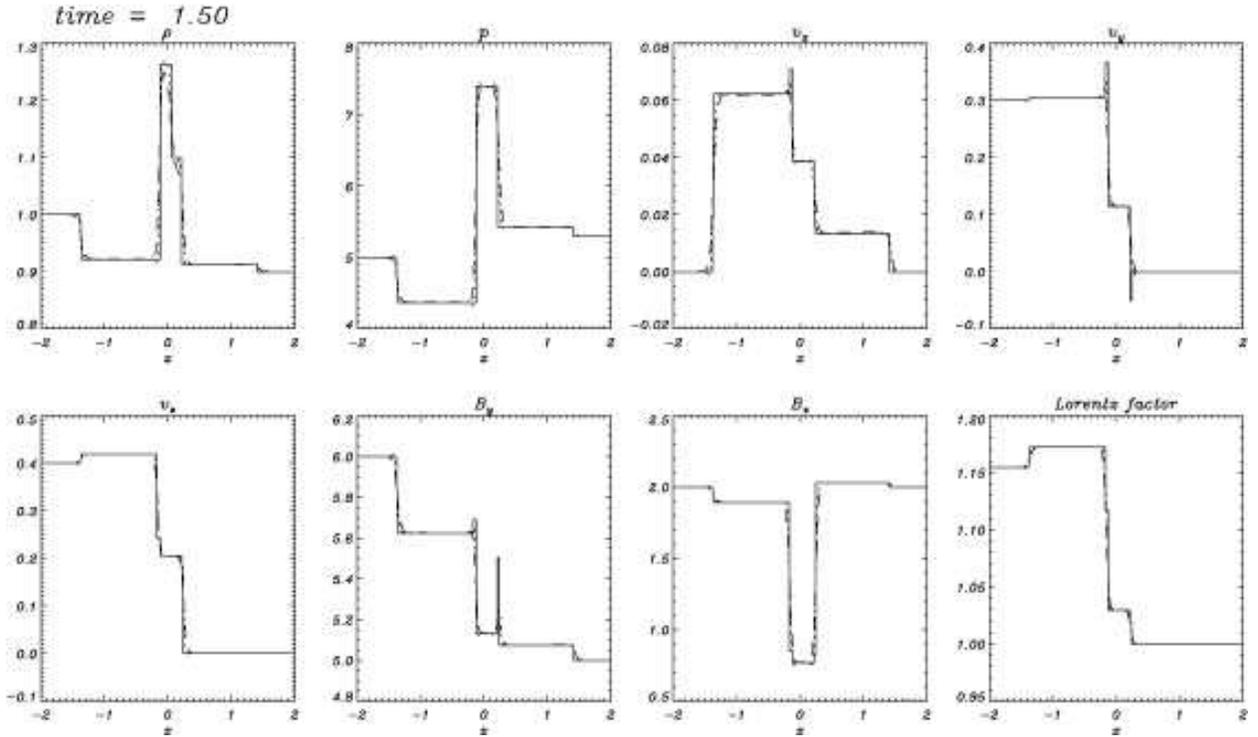} \caption{ Simulation results of Generic
Alfv\'{e}n test at time $t=1.5$ using the MC slope limiter (dotted
lines) and the minmod slope limiter (dashed lines) reconstructions.
The solid lines are the exact solution. The results are composed of
a left-going fast rarefaction,  a left-going Alfv\'{e}n
discontinuity, a left-going slow shock, a contact discontinuity, a
right-going slow shock, a right-going Alfv\'{e}n discontinuity, and
a right-going fast shock. \label{f10a}}
\end{figure}

\begin{figure}[ht]
\plotone{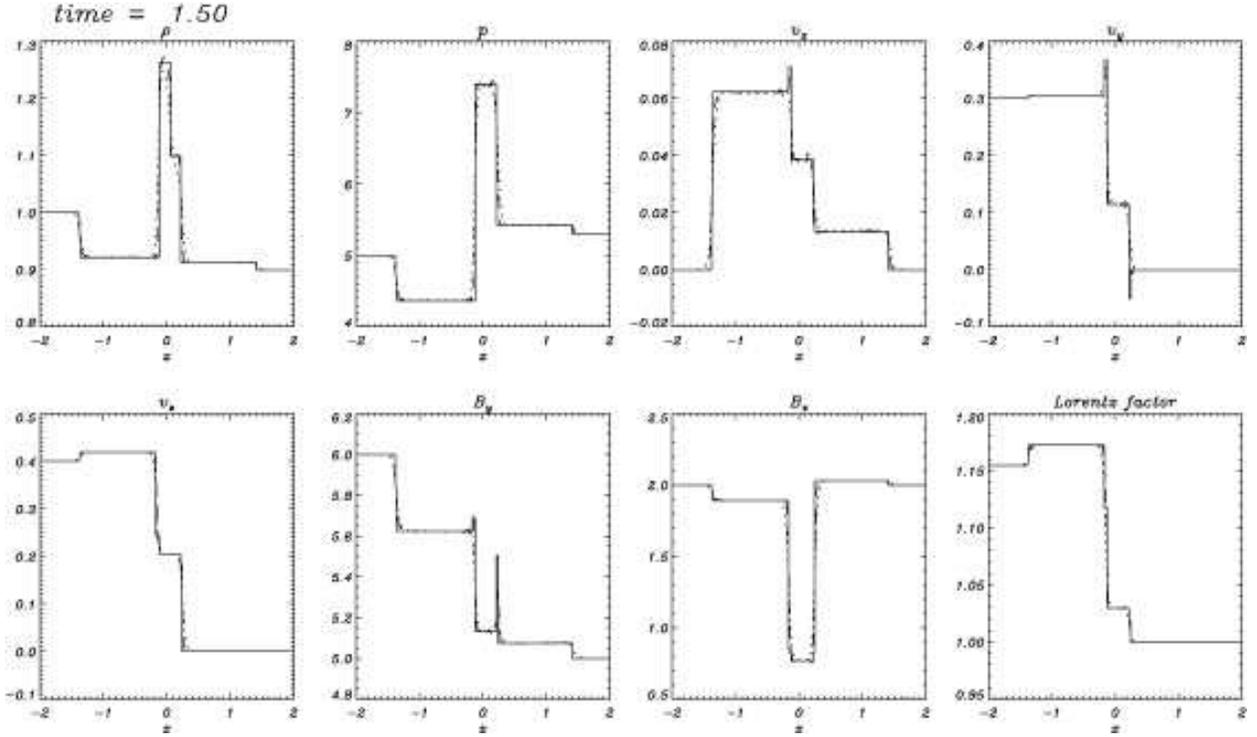} \caption{ Simulation results of Generic
Alfv\'{e}n test at time $t=1.5$ using the CENO (dotted lines) and
the PPM (dashed lines) reconstructions.  The solid lines are the
exact solution. The results are composed of a left-going fast
rarefaction, a left-going Alfv\'{e}n discontinuity, a left-going
slow shock, a contact discontinuity, a right-going slow shock, a
right-going Alfv\'{e}n discontinuity, and  a right-going fast shock.
\label{f10b}}
\end{figure}

The results of the simulations are shown in Fig \ref{f3a}
-\ref{f10b}. All results show  good agreement with the exact
solutions. However, some small discontinuities and large shocks
cannot be resolved exactly. This means that we need more
computational zones to resolve all small discontinuities and large
shocks exactly. Generally, the minmod slope limiter and CENO
reconstructions are more diffusive than the MC slope limiter and PPM
reconstructions because of the properties of the minmod function. On
the other hand, although the MC slope limiter and PPM
reconstructions can resolve sharp discontinuities well, some small
oscillations are seen at the discontinuities. The PPM reconstruction
is the most accurate scheme to detect the discontinuities.

Our previous GRMHD code of Koide (2003) could not handle some
extreme cases of relativistic MHD shock-tube tests shown in Table
\ref{table1} such as Kommissarov: Shock-Tube test1, Balsara Test2
and Balsara Test3 for large discontinuities of the pressure and
magnetic field, Kommissarov: Collision Test and Balsara Test3 for
the highly relativistic flow even using different recovery methods
such as Noble 2D method. However the new GRMHD code successfully
handles the relativistic MHD shock-tube tests in Table \ref{table1}.
Therefore the new GRMHD code has substantial improvements over our
previous GRMHD code (e.g., Koide 2003) and can operate in a regime
with large discontinuities of physical quantities (4 order
difference of pressure in Komissarov: Collision Test and Balsara
Test3), strong magnetic field ($\beta < 0.004$ and $\sigma > 570$ in
Balsara Test 3, where $\beta=p_{gas}/p_{mag}$, $p_{mag}= b^{2}/2$,
and $\sigma=|b^{2}|/\rho$) and highly relativistic flow ($\gamma >
22$ in Balsara Test 4). The limitation to handle the regimes of high
Lorentz factor and of highly magnetized depends on the schemes to
solve the GRMHD equations.

\subsection{Magnetized Bondi flow}

Next we check the code to verify it numerically maintains the
time-dependent system of equations describing the stationary,
spherically symmetric accretion of a perfect fluid onto a
Schwarzschild black hole in the presence of a radial magnetic field.
Spherically symmetric accretion (Bondi flow) onto the fixed
background of the Schwarzschild black hole has an analytic solution
(e.g., Shapiro \& Teukolsky 1983) that can be compared with the
results of our code. This test has been used by several authors (De
Villiers \& Hawley 2003; Gammie et al. 2003; Duez et al. 2005;
Shibata \& Sekiguchi 2005; Ant\'{o}n et al. 2006) in the validation
of their GRMHD codes.

The initial setup consists of a perfect fluid which satisfies a
$\Gamma$-law EOS with $\Gamma=4/3$. The analytic solution is
calculated in a manner similar to Koide et al. (1999) with constant
parameter $H=\rho_{0} h_{0}=1.3$ and $\rho_{0} = 1.0$.
The critical point of free-falling flow is located at $r_{\rm S}=3.0$.
The radial magnetic field component is chosen to satisfy the divergence-free
condition. Its strength is determined by the parameter $B_{0}$ at
the critical radius of the flow. These initial conditions are
evolved in time in a uniform radial gird covering the region $1.1
r_{\rm S} \le r \le 20 r_{\rm S}$.

\begin{figure}[ht]
\epsscale{0.8}
\plotone{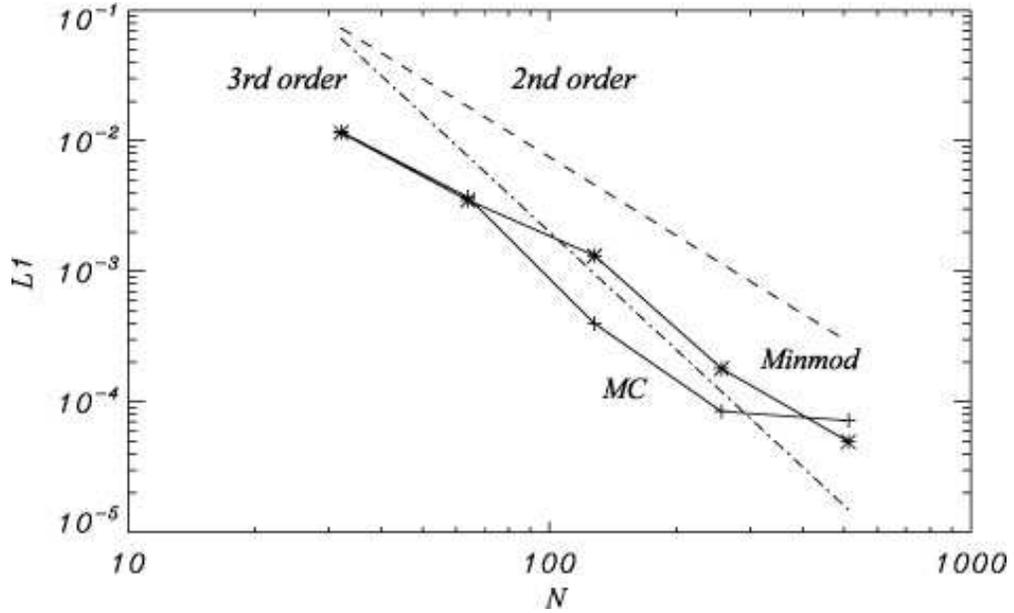}
\caption{ $L_{1}$ norm of the error in $v^{x}$ for a magnetized
Bondi accretion flow as a function of computational zone number $N$ for
the MC slope limiter
(plus) and the minmod slope limiter (asterisk) reconstructions.
The straight lines show the slope expected for second-order
convergence (dashed line) and third-order convergence (dash-dotted
line). \label{f2}}
\end{figure}

Figure \ref{f2} shows the $L_{1}$ norm of the radial velocity
$v^{x}$  ($Q=v^{x}$ in Eq. (\ref{eq4-1})) of the difference between
the exact solution and the final state ($t_{end}=50 \tau_{s}$,
$\tau_{s} = r_{\rm S}/c$ ) in the $B_{0} = 0.001 \sqrt{\rho_{0} c^{2}}$ case.
The calculation of $L_{1}$ norm is taken in the region $1.5 r_{\rm S}
\le r \le 18 r_{\rm S}$ to exclude boundary effects. The MC slope
limiter and minmod slope limiter reconstructions show that the
global order of convergence of the code is second order.
The MC slope limiter reconstruction is more accurate than the minmod
slope limiter reconstruction.

\section{Summary and Conclusions}

We have developed a new three-dimensional GRMHD code, RAISHIN, by
using a conservative high resolution shock-capturing scheme. The
numerical fluxes are calculated using the HLL approximate Riemann
solver scheme. The flux-interpolated, constrained transport scheme
is used to maintain a divergence-free magnetic field. Several
reconstruction and time advance schemes can be chosen for the
numerical accuracy and computational resources.

We have described several test problems in both special and general
relativity. They have shown significant improvements over our
previous GRMHD code (Koide 2003). Our new GRMHD code can perform in
the regimes of high Lorentz factor ($ \gamma >20 $) and
high magnetic field ($\sigma > 550$), and in the presence of
large discontinuity of density, pressure and magnetic field.
We have compared the results of several reconstruction
schemes. The code is second-order accurate even when we use the higher
order reconstruction schemes such as CENO and PPM. Nevertheless,
higher-order reconstruction schemes can provide more accurate
results for some applications. The PPM reconstruction scheme allows
the well-resolved detection of sharp discontinuities.
We found the limitation to handle the regimes of high Lorentz factor
and of highly magnetized depends on the schemes to solve the GRMHD
equations.

We have performed several simulations of non-rotating and rotating
black hole systems with a thin accretion disk (Mizuno et al. 2006).
The simulation results show the formation of jets driven by the
Lorentz force and the gas pressure. It appears that the rotating
black hole creates an additional, faster, and more collimated inner
outflow beside an outflow generated by the rotating accretion disk
in the non-rotating black hole. Thus, kinematic jet structure could
be a sensitive function of the black hole rotation.

Our new GRMHD code has proven to be accurate in second order and has
successfully passed all applied tests including highly relativistic
cases , and highly magnetized cases
in both special and general relativity. We plan to apply this code to
a number of high-energy
astrophysical phenomena involving highly relativistic flows or
compact objects with strong gravitational fields and magnetic
fields.

\acknowledgments

 Y. M. is a NASA Postdoctoral Program fellow at NASA
Marshall Space Flight Center. He thanks the help of B.
Giacomazzo for providing their calculation code to get exact the
solution of RMHD Riemann problems. He also thanks G. Richardson,
D. Hartmann, C. Fendt, and M. Camenzind for useful discussions. K. N.
is partially supported by
the National Science Foundation awards ATM-0100997, INT-9981508, and
AST-0506719, and the National Aeronautic and Space Administration
award NASA-INTEG04-0000-0046 to the Univ of Alabama in Huntsville.
P.H. acknowledges partial support by a National Space Science and Technology
(NSSTC/NASA) cooperative agreement NCC8-256 and NSF awards AST-0506666.
The simulations have been performed on
IBM p690 at the National Center for Supercomputing Applications
(NCSA) which is supported by the NSF and  Altix3700 BX2 at YITP in
Kyoto University.

\newpage

\begin{deluxetable}{llcccccccc}
\tabletypesize{\small}
\tablecolumns{10}
\tablewidth{0pc}
\setlength{\tabcolsep}{0.04in}
\tablecaption{Relativistic MHD Shock-Tube Tests \label{table1}}
\tablehead{
\colhead{Test Type} &  & \colhead{$\rho$} & \colhead{$p$} & \colhead{$v^{x}$} &
\colhead{$v^{y}$} & \colhead{$v^{z}$} & \colhead{$B^{x}$} & \colhead{$B^{y}$} &
\colhead{$B^{z}$} }
\startdata
\textbf{Komissarov: Shock Tube Test1} & {\itshape left state\/} & 1.0 & 1000.0
& 0.0 & 0.0 & 0.0 & 1.0 & 0.0 & 0.0 \\
$\Gamma=4/3$, $t_{\mathrm{final}}=1.0$  & {\itshape right state\/} & 0.1 & 1.0
& 0.0 & 0.0 & 0.0 & 1.0 & 0.0 & 0.0 \\
 \hline
\textbf{Komissarov: Collision Test}  & {\itshape left state\/} & 1.0 & 1.0 & $5/
\sqrt{26}$ & 0.0 & 0.0 & 10.0 & 10.0 & 0.0 \\
$\Gamma=4/3$,  $t_{\mathrm{final}}=1.2$ & {\itshape right state\/} & 1.0 & 1.0 & $-5/
\sqrt{26}$ & 0.0 & 0.0 & 10.0 & -10.0 & 0.0 \\
 \hline
\textbf{Barsara Test1 (Brio \& Wo)} & {\itshape left state\/} & 1.000 & 1.0 & 0.0
& 0.0 & 0.0 & 0.5 & 1.0 & 0.0 \\
$\Gamma=2$, $t_{\mathrm{final}}=0.4$ & {\itshape right state\/} & 0.125 & 0.1 & 0.0
& 0.0 & 0.0 & 0.5 & -1.0 & 0.0 \\
 \hline
\textbf{Barsara Test2} & {\itshape left state\/} & 1.0 & 30.0 & 0.0 & 0.0 & 0.0
& 5.0 & 6.0 & 6.0 \\
 $\Gamma=5/3$, $t_{\mathrm{final}}=0.4$ & {\itshape right state\/} & 1.0 & 1.0
 & 0.0 & 0.0 & 0.0 & 5.0 & 0.7 & 0.7 \\
 \hline
\textbf{Barsara Test3} & {\itshape left state\/} & 1.0 & 1000.0 & 0.0 & 0.0 & 0.0
& 10.0 & 7.0 & 7.0 \\
$\Gamma=5/3$, $t_{\mathrm{final}}=0.4$ & {\itshape right state\/} & 1.0 & 0.1 & 0.0
& 0.0 & 0.0 & 10.0 & 0.7 & 0.7 \\
 \hline
 \textbf{Barsara Test4} & {\itshape left state\/} & 1.0 & 0.1 & 0.999 & 0.0 & 0.0
 & 10.0 & 7.0 & 7.0 \\
$\Gamma=5/3$, $t_{\mathrm{final}}=0.4$ & {\itshape right state\/} & 1.0 & 0.1
& -0.999 & 0.0 & 0.0 & 10.0 & -7.0 & -7.0 \\
 \hline
\textbf{Barsara Test5} & {\itshape left state\/} & 1.08 & 0.95 & 0.40 & 0.3 & 0.2
& 2.0 & 0.3 & 0.3 \\
$\Gamma=5/3$, $t_{\mathrm{final}}=0.5$ & {\itshape right state\/} & 1.00 & 1.0
& -0.45 & -0.2 & 0.2 & 2.0 & -0.7 & 0.5 \\
 \hline
\textbf{Generic Alfv\'{e}n Test} & {\itshape left state\/} & 1.0 & 5.0 & 0.0 & 0.3
& 0.4 & 1.0 & 6.0 & 2.0 \\
$\Gamma=5/3$, $t_{\mathrm{final}}=1.5$ & {\itshape right state\/} & 0.9 & 5.3 & 0.0
& 0.0 & 0.0 & 1.0 & 5.0 & 2.0 \\

\enddata
\tablecomments{Initial conditions for the relativistic shock-tube tests.}
\end{deluxetable}

\newpage

\newpage

\newpage

\newpage

\newpage

\newpage

\newpage

\newpage

\newpage

\newpage

\newpage

\newpage

\newpage

\newpage

\newpage

\newpage

\newpage

\newpage

\end{document}